\documentclass[aip,jap,reprint,10pt,showpacs,altaffillsymbol,superscriptaddress,longbibliography]{revtex4-1}
\usepackage{graphicx}
\usepackage{amsfonts}
\usepackage{amsmath}
\usepackage{natbib}
\usepackage{xcolor}
\pdfoutput=1

\begin{document}

\title{Evidence for universality of tunable-barrier electron pumps}

\author{Stephen~P.~Giblin}
\affiliation{National Physical Laboratory, Hampton Road, Teddington, Middlesex TW11 0LW, United Kingdom}
\author{Akira Fujiwara}
\affiliation{NTT Basic Research Laboratories, NTT Corporation, 3-1 Morinosato Wakamiya, Atsugi, Kanagawa 243-0198, Japan}
\author{Gento Yamahata}
\affiliation{NTT Basic Research Laboratories, NTT Corporation, 3-1 Morinosato Wakamiya, Atsugi, Kanagawa 243-0198, Japan}
\author{Myung-Ho Bae}
\author{Nam Kim}
\affiliation{Korea Research Institute of Standards and Science, Daejeon 34113, Republic of Korea}
\author{Alessandro Rossi}
\affiliation{Cavendish Laboratory, University of Cambridge, JJ Thomson Avenue, Cambridge CB30HE, United Kingdom}
\author{Mikko M\"ott\"onen}
\affiliation{QCD Labs, COMP Centre of Excellence, Department of Applied Physics, Aalto University, 00076 AALTO, Finland}
\author{Masaya Kataoka}
\affiliation{National Physical Laboratory, Hampton Road, Teddington, Middlesex TW11 0LW, United Kingdom}

\email[stephen.giblin@npl.co.uk]{Your e-mail address}

\date{\today}

\begin{abstract}
We review recent precision measurements on semiconductor tunable-barrier electron pumps operating in a ratchet mode. Seven studies on five different designs of pumps have reported measurements of the pump current with relative total uncertainties around $10^{-6}$ or less. Combined with theoretical models of electron capture by the pumps, these experimental data exhibits encouraging evidence that the pumps operate according to a universal mechanism, independent of the details of device design. Evidence for robustness of the pump current against changes in the control parameters is at a more preliminary stage, but also encouraging, with two studies reporting robustness of the pump current against three or more parameters in the range of $\sim\!5 \times 10^{-7}$ to $\sim\!2 \times 10^{-6}$. This review highlights the need for an agreed protocol for tuning the electron pump for optimal operation, as well as more rigorous evaluations of the robustness in a wide range of pump designs.

\end{abstract}

\pacs{1234}

\maketitle

\section{\label{IntroSec} Introduction}

In the mid-to-late 1980s, advances in nano-fabrication techniques made controlled transport of single charges in solid-state devices possible. The first proposal to make a current standard by linking current to the elementary charge and frequency also dates from this period \cite{likharev1985theory}. Subsequently, single-electron turnstiles (which require a bias voltage)\cite{geerligs1990frequency} and pumps (which do not)\cite{pothier1992single} were demonstrated in metal-oxide nano-structures, transporting electrons one at a time at frequencies up to about $10$~MHz to generate currents of a few pA. Following major research projects principally at NIST (USA) and PTB (Germany), multi-junction metal oxide pumps demonstrated sub part-per-million (ppm) electron transfer accuracy \cite{keller1996accuracy,camarota2012electron}, and were operated as prototype capacitance standards \cite{keller1999capacitance,camarota2012electron}. Ultimately, despite these promising results, the metal-oxide pumps were limited to pA-level currents by the fixed time constants of the tunnel barriers, and directly scaling the pump currents, for example using cryogenic current comparators (CCCs)\cite{steck2008characterization} was found to be challenging.


Semiconductor systems in principle provide a more flexible platform for investigating single-electron effects in the solid state, because tunnel barriers can be formed electrostatically using voltages applied to gates\cite{kouwenhoven1997electron}, and can therefore be adjusted unlike the fixed barriers defined by oxide layers in metal-oxide systems. Gallium Arsenide (GaAs), being piezoelectric, also supports surface acoustic waves (SAWs), and some promising results were achieved using SAWs at GHz frequencies to transport electrons through a potential barrier defined using gates\cite{shilton1996high,janssen2000accuracy,fletcher2003quantized}. Currents of up to $\sim\!0.5$~nA were demonstrated with an accuracy\cite{janssen2000accuracy} of $\sim\!10^{-4}$, although the SAW approach was later abandoned in favour of modulating the barrier gates. A two-gate turnstile was first demonstrated in 1991 \cite{kouwenhoven1991quantized}, but the breakthrough results were obtained 15 years later, roughly simultaneously by groups at Cambridge University and NPL (UK)\cite{blumenthal2007gigahertz}, NTT Basic Research Laboratories (Japan)\cite{fujiwara2008nanoampere}, and PTB (Germany)\cite{kaestner2008single}. These results showed that electrons could be pumped in a ratchet mode at GHz frequencies using (in two of the studies)\cite{kaestner2008single,fujiwara2008nanoampere} only one high-frequency control gate, generating currents $I_{\text{P}}$ up to $\sim 200$~pA. Stimulated by these promising results, metrological investigations of the pump accuracy were undertaken at NPL and PTB. To date, seven studies \cite{giblin2012towards,bae2015precision,stein2015validation,yamahata2016gigahertz,stein2016robustness,giblin2017robust,zhao2017thermal} have reported comparison of electron pump currents with reference currents derived indirectly from the quantum Hall effect (QHE) and the Josephson voltage standard (JVS), with relative combined uncertainties of $10^{-6}$ or less. These studies reported measurements of five types of tunable-barrier semiconductor pumps fabricated from silicon\cite{yamahata2016gigahertz,zhao2017thermal} and GaAs \cite{giblin2012towards,bae2015precision,stein2015validation,stein2016robustness,giblin2017robust} with widely differing channel and gate geometries, all tuned to pump one electron per cycle. The fact that all these pumps could be operated successfully with a part-per-million accuracy suggests that they are transferring one elementary charge per cycle independently of the design details. Additionally, two of the studies\cite{stein2016robustness,giblin2017robust} demonstrated significant robustness of the pump current to changes in the control parameters of the pumps.

Universality is a key concept underlying the use of quantum standards, and is already familiar to the electrical metrology community through the widespread adoption of the QHE and JVS as standards of resistance and voltage, respectively. It means that the operation of the standard is based on fundamental principles which are manifested in the same way in severely different physical realisations of the standard. Two types of experimental data support the hypothesis of universality: \textit{robustness} of the realised parameter (in the case of electron pumps, pumped current) as the tuning parameters of the device are varied, and \textit{agreement} between the resulting robust currents generated by different devices. The acceptance of the QHE and JVS as primary standards was driven by a three-way interplay of robustness, agreement, and the third ingredient, rigorous theoretical analysis showing the fundamental nature of the underlying principle. To give examples for the case of the QHE, a key experimental study of Ref. \onlinecite{jeckelmann1997high} demonstrates agreement and robustness (as a function of gate voltage for Si MOSFET Hall devices and many other parameters including temperature and step number), while more recently, relative agreement at a level below $10^{-10}$ was demonstrated between the QHE in graphene and GaAs Hall devices\cite{janssen2012precision}. Underpinning these empirical findings lies a body of theoretical knowledge which relates the QHE to topological invariants which describe broad classes of solid state systems \cite{thouless1994topological}.

In this paper, we review the experimental data for robustness and agreement of tunable-barrier electron pumps, as well as the theoretical underpinning of their operation, and consider to what extent there is evidence for universality. We also consider what further work is needed on tunable barrier pumps before they can be considered primary standards, and hence play a routine role in primary electrical metrology. This is not intended to be a comprehensive review of single-electron pumps; we do not discuss devices such as the hybrid turnstile \cite{pekola2007hybrid,maisi2009parallel}, which have not yet demonstrated metrological accuracy but which may do so in the future. For a comprehensive review of the physics and technology of all single-electron transfer devices, the reader is referred to Ref. \onlinecite{pekola2013single}. Other reviews have been published covering quantum current standards\cite{kaneko2016review}, more specifically semiconductor electron pumps\cite{kaestner2015non} and very recently, the role of electron pumps in the context of the revision of the SI system \cite{scherer2019single}. In this review, we are implicitly considering a future scenario in which electron pumps (or parallel arrays of pumps) are used as primary current standards following a setup procedure still to be agreed on, but analogous to that already in use for the QHE\cite{delahaye2003revised}. Another powerful paradigm, outside the scope of this review, also exists for the metrological use of electron pumps: the self-referenced current standard based on real-time counting of errors made by a series array of pumps\cite{fricke2013counting,fricke2014self,scherer2019single}. This method requires less rigorous characterisation of the pumps making up the array, because in principle all electron transfer errors can be accounted for, but it has so far demonstrated modest accuracy at current levels well below $1$~pA\cite{fricke2014self}.

The paper is structured as follows: In section \ref{BasicSec}, we describe the ratchet mode of operation of the tunable-barrier pump, and in section \ref{FabSec}, we briefly describe the fabrication technology which has enabled ratchet-mode devices to be realised in silicon and GaAs material systems. In section \ref{TheorySec}, we discuss in some detail theoretical work on the ratchet-mode devices. In section \ref{TuningSec}, we describe how the devices are tuned to achieve the high-accuracy pumping regime, and in section \ref{MethSec}, we discuss the techniques used to compare the pump current to known reference currents. Finally in section \ref{ResultsSec}, we review the experimental evidence for robustness and agreement from the seven precision studies \cite{giblin2012towards,bae2015precision,stein2015validation,yamahata2016gigahertz,stein2016robustness,giblin2017robust,zhao2017thermal} under consideration.

\section{\label{BasicSec} Basic operation of the tunable-barrier pump}

\begin{figure}
\includegraphics[width=8.5cm]{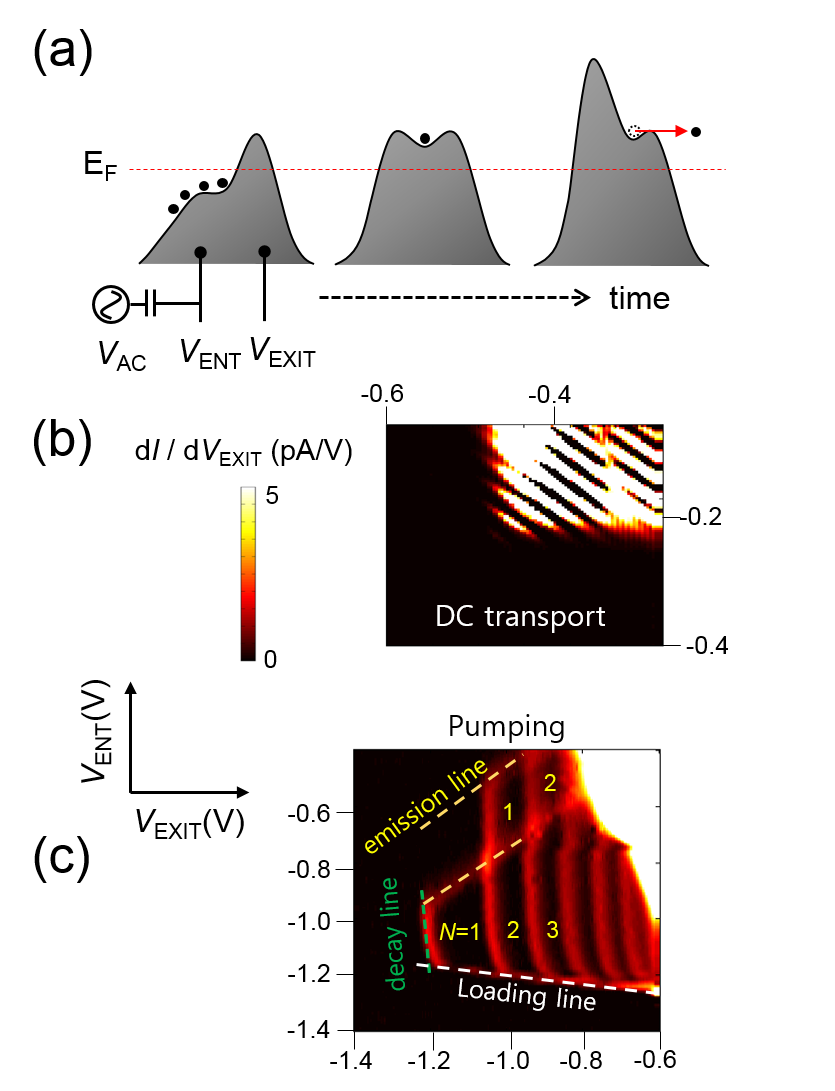}
\caption{\label{Fig1}\textsf{(a): Schematic potential diagrams illustrating a sequence of pumping a single electron from left to right through a tunable-barrier pump. Two DC voltages $V_{\text{ENT}}$ and $V_{\text{EXIT}}$ define two potential barriers, and the AC voltage $V_{\text{AC}}$ is also applied to the entrance gate. The Fermi energy is denoted $E_{\text{F}}$. (b): Colour-map plot of $\text{d}I / \text{d}V_{\text{EXIT}}$ for an electron pump (with $\sim 1$~meV Coulomb gap) in the static QD regime, as a function of entrance and exit gate voltages. $V_{\text{AC}}=0$. (c): Colour-map plot of the same device in the pumping regime: $\text{d}I_{\text{P}} / \text{d}V_{\text{EXIT}}$ data measured at $f = 100$~MHz with $V_{\text{SD}}=0$. The applied source-drain bias voltage is denoted $V_{\text{SD}}$, and $N$ denotes the average number of electrons pumped in each cycle. Plots (b) and (c) share the same colour bar scale.}}
\end{figure}

The three sequential energy diagrams of figure 1(a) depict the process of single electron pumping in the ratchet mode; loading, trapping, and unloading. These diagrams pre-suppose that electrons are confined into a one-dimensional channel which is crossed by two electrostatic gates, so that a one-dimensional cut along the device axis is sufficient to illustrate the pump operation. We use the term `ratchet mode' because an alternating voltage applied to a single gate induces a direct (DC) current. The precision metrological results considered in this paper were all obtained with an alternating voltage applied to only one gate, although there is some evidence\cite{stein2015validation} that driving both gates with phase-shifted signals may be advantageous at high frequencies. Other modes of electron transfer, distinct from the ratchet mode, have been demonstrated using two\cite{kouwenhoven1991quantized,jehl2013hybrid}, or three\cite{tanttu2016three} driven gates, but these have not so far demonstrated metrological accuracy. The two gates are referred to as the entrance and exit gates, and the DC voltages applied to them are denoted by $V_{\text{ENT}}$ and $V_{\text{EXIT}}$. Using the gates to raise the potential barriers creates a quantum dot (QD) in between the gates \cite{kouwenhoven1997electron}. An alternating (AC) voltage, $V_{\text{AC}}$, is added to the entrance gate to drive the pumping. In order to realize the single-parameter ratchet mode, the strength of the cross-coupling between the entrance gate and the QD potential should be optimal. This optimization is discussed in more detail in section \ref{TheorySec}. For the pumping operation, the QD potential must be lifted above the Fermi level by the cross-coupling at the capturing stage in the pumping sequence as illustrated in the middle panel of figure 1(a). Thus far no general rule has been formulated describing how to design a single-parameter ratchet pump device\cite{zimmerman2018one-gate}, and each research group has developed its own prototypes of the pump devices empirically.

In the parameter space spanned by $V_{\text{ENT}}$ and $V_{\text{EXIT}}$, the pumping zone is located in the region where the channel is non-conducting due to the large potential barriers formed by the two gates. Figure \ref{Fig1} (b) shows the conductance reducing to zero as the gate voltages are made more negative. Coulomb blockade oscillations are also visible in this device. Figure \ref{Fig1} (c) shows a characteristic pump map, the fingerprint of ratchet mode pumping in which the number of pumped electrons in each cycle, $N$, is demarcated by clear lines. For instance, $V_{\text{EXIT}}$ determines the number of electrons captured while $V_{\text{ENT}}$ determines the number of electrons not emitted at the un-loading stage \cite{kaestner2015non}. As long as the electrons stay in their ground states, pumping does not occur outside of the event lines of loading, decay, and emission indicated in figure 1(b). All of the high-precision studies discussed in this review utilize the $N=1$ plateau.


\section{\label{FabSec} Device Technology}
\begin{figure*}
\includegraphics[width=17 cm]{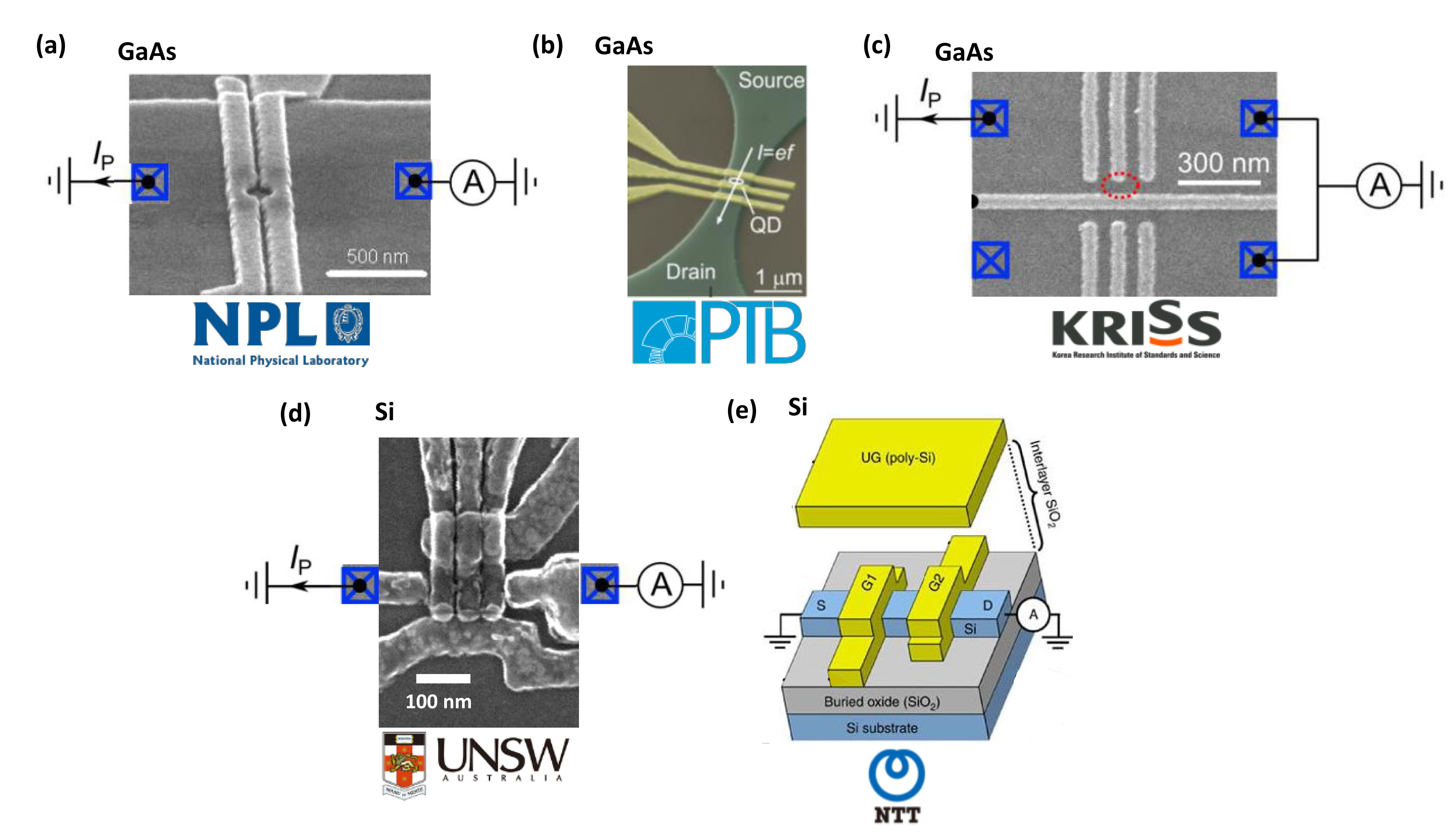}
\caption {Images illustrating the five single-electron pump devices discussed in this article. The semiconductor of choice and the manufacturing institution are indicated for each device. (a): SEM image (from figure 1 of Ref. \onlinecite{giblin2012towards}) of a GaAs pump with a $2$~$\mu$m-wide etched channel and a cut-out in the middle of the gates for forming the QD. (b): False-coloured SEM image (from figure 1 of Ref. \onlinecite{stein2016robustness}) of a GaAs pump with a narrower etched channel and straight gates, as used in Refs. \onlinecite{stein2015validation,stein2016robustness}. (c): SEM image (from figure 1 of Ref. \onlinecite{giblin2017robust}) of a GaAs pump without an etched channel, and with the QD (approximately indicated by the dotted red circle) defined entirely using surface gates, as used in Refs. \onlinecite{bae2015precision,giblin2017robust}. (d): SEM image (from figure 1 of Ref. \onlinecite{rossi2014accurate}) of a silicon pump, as used in Ref. \onlinecite{zhao2017thermal}. (e): Schematic diagram (from Ref. \onlinecite{yamahata2014gigahertz}) of the nano-wire MOSFET device used in Ref. \onlinecite{yamahata2016gigahertz}.}
\label{devices}
\end{figure*}

In this section, we discuss the single-electron pump device technology developed by five institutions: KRISS (South Korea), NPL in collaboration with Cambridge University (UK), NTT (Japan), PTB (Germany), and UNSW (Australia). As mentioned earlier, the main reason for the success of semiconductor electron pumps over competing technologies is the possibility to electrostatically control their tunnel barriers. At the device level, this feature is typically implemented with gate electrodes that employ the field effect to locally regulate the electron density.

A two-dimensional electron layer accumulated at the interface between different materials is used as the active layer where single-charge transport occurs. Depending on the semiconductor of choice, its realisation may differ. For example, as indicated in figure~\ref{devices} (a)--(c), the pumps made by KRISS, NPL and PTB are based on GaAs/AlGaAs heterostructures. Doping modulation in these two materials produce sufficient conduction band bending at the interface to form a potential well for electrons. The electron layer is typically located between 60 and 100 nm below the wafer surface. The pumps made by NTT and UNSW are based on silicon metal-oxide-semiconductor (MOS) accumulation-mode transistors, as shown in figure~\ref{devices} (d)--(e). In this case, the conduction band of a near-intrinsic silicon substrate can be locally bent by positive gate voltages, resulting in the formation of an accumulation layer at the interface between Si and thermally-grown SiO$_2$. The electron layer is typically found between 6 nm (UNSW) and 30 nm (NTT) below the gates depending on the thickness of the gate dielectric.

Stemming from the different origin of the device active layer, the type of action required for the gates to create tunable tunnel barriers changes. In particular, in all GaAs devices the gates have to locally deplete the electron layer. This is achieved by realizing metal electrodes on the wafer surface via standard electron-beam (e-beam) lithography and deposition techniques. The resulting metal/semiconductor Schottky junction is reverse biased by applying negative gate voltages, leading to the desired amount of electron depletion. Similar bias conditions are needed for the MOS pumps made by NTT. This is because a large poly-Si top gate is used to accumulate the electron layer in the silicon-on-insulator channel, whereas two 50-nm-wide lower gates have the function to locally deplete it. Electrical insulation between the upper and lower gate layer is achieved by chemical vapour deposition of an oxide. The silicon pumps made by UNSW have e-beam-defined Al gates arranged in a 3-layer stack and inter-layer insulation is obtained using thermally grown Al oxide. In contrast to the NTT devices, there is no large gate inducing an electron layer over the entire conductive channel. Hence, multiple densely packed gates are all positively biased leading to a continuous electron accumulation layer between source and drain. In this configuration, it is sufficient to apply slightly less positive voltages to some gates to define and control the tunnel barriers.

Tight charge confinement is paramount to generate accurate currents with any of these electron pumps. The gate-controlled tunnel barriers provide confinement in the longitudinal direction, i.e. the direction of transport. However, transverse confinement is also necessary and the way this is achieved differs in the discussed device technologies. For example, the pumps from NTT and PTB physically confine electrons in the transverse direction by etching the conductive channel. This leads to the formation of quasi-1 dimensional semiconducting wires that in the gated regions have approximate widths of 15 nm (NTT) and 680 nm (PTB). The 1D wire is realised using e-beam lithography and wet or dry\cite{fujiwara1998suppression} etching techniques. By contrast, the pumps from KRISS and UNSW have planar bulk substrates and, hence, transverse confinement is achieved electrostatically employing dedicated gates. Despite the additional device complexity, these pumps provide enhanced tunability of the QD size, which has proved beneficial for high-accuracy operation. Finally, the NPL devices use a combination of physical and electrostatic confinement. Although a 1500-nm-wide wire is etched in the semiconductor, most of the transverse confinement is achieved by a tailored design of the gate electrodes, as shown in figure~\ref{devices}(a).

\section{\label{TheorySec} Theory of the tunable-barrier pump}

The electron pump aims to exploit the fundamental quantisation of charge in order to generate an electric current by moving electrons one at a time. When discussing the theory in this section, however, we do not focus on the underlying phenomenon of charge quantisation in solid-state devices, but on the theory describing how the ratchet-mode tunable-barrier pumps, as a specific class of devices, transfer charges. We take the underlying charge quantisation for granted, but for completeness we note that in an earlier precision study on the adiabatic electron pump\cite{keller1999capacitance}, the measurement of the current, combined with a separate single-electron error counting measurement \cite{keller1996accuracy} was interpreted as closing the `metrological triangle'\cite{keller2008current} at a relative uncertainty of $10^{-6}$. 

As described below, the theory of the tunable-barrier pump involves detailed considerations of the time-dependent tunnel rates into and out of the QD, which itself has a time-dependent electrostatic potential. We show in sections \ref{TheorySec} B--D that the current as a function of exit gate for many pumps can be explained by the `decay cascade' model, in which electrons tunnel out of the dynamically forming QD, and furthermore (sub-section E) that similar data on some pumps can be explained by a `thermal capture' model in which the QD exchanges electrons with the lead many times during its formation. We next show in section \ref{TheorySec} F that the main features of $I_{\text{P}} (V_{\text{EXIT}})$ data in both the decay cascade and thermal capture models can be reproduced using a simple monte-carlo simulation, and we briefly discuss in section \ref{TheorySec} G the upper frequency limit of the pump operation which is largely an unsolved problem.

\subsection{Theoretical background}
Unfortunately, the tunable-barrier pump is theoretically rather intractable and no exact microscopic theory exists to explain its operation. To understand why this is the case, it is instructive to first briefly consider the metal-oxide pump in its adiabatic limit. These pumps operate in a regime defined by three inequalities: $k_{\text{B}}T \ll E_{\text{C}} = \frac{e^2}{C_{\Sigma}}$, $R \gg R_{\text{K}}$ and $f \ll \Gamma$, where $k_{\text{B}}$ is the Boltzmann constant, $T$ is temperature, $E_{\text{C}}$ and $C_{\Sigma}$ are the Coulumb energy and the capacitance of the pump islands, respectively, $R$ is the resistance of the tunnel barriers, $R_{\text{K}}$ is the resistance quantum, $f$ is the pump frequency and $\Gamma$ is the tunnel rate for electrons through the junctions. The first inequality guarantees that the number of electrons on each island is stable against thermal fluctuations, the second implies that the number of electrons on each island is well defined and stable against quantum fluctuations, and the third states that the energy of each island is manipulated (using gate voltages) sufficiently slowly that the energetically favorable configuration of the pump is always attained at each point in the pumping cycle. It is this third inequality that limits the metal-oxide pumps to currents of order $1$~pA. The quantities $\Gamma$, $R$ and $E_{\text{C}}$ are inter-related, which imposes additional constraints. Increasing $\Gamma$, for example by making the tunnel barrier with a thinner oxide layer, will reduce $R$ and also slightly reduce $E_{\text{C}}$ by increasing the barrier capacitance. Nevertheless, these assumptions allow a theory of the pump operation to be constructed using a master-equation formalism, which allows, for example, quantitative predictions of how the pump error increases with increasing frequency or bias voltage \cite{martinis1994metrological}. Extensions to the theory which include absorption and emission of energy from the environment have allowed quantitative predictions of photon-assisted tunneling errors \cite{covington2000photon}, study of the effect of engineered high-impedance environments\cite{zorin2000coulomb,lotkhov2001operation} and even helped solve practical problems like the design of cryogenic electrical filters \cite{zorin1995thermocoax}.

The tunable-barrier pump achieves its high pumping rate precisely because the barrier resistance is very small at the points in the pumping cycle when electrons are transferred between the leads and the pump, i.e, it may strongly violate all three of the above inequalities. Two key insights have allowed theoretical progress: firstly, only the loading phase matters, because sufficiently large amplitude $V_{\text{AC}}$, can eject the loaded electrons to the drain with unity probability \cite{kaestner2008robust,miyamoto2010resonant}. This insight has important practical consequences, because a special drive waveform from an arbitrary waveform generator (AWG) can be used to optimise the capture phase of the pump cycle \cite{giblin2012towards,stein2015validation,stein2016robustness}. Secondly, the charge quantisation during the loading phase is not a static process defined by a single energy scale as in metallic pumps: it is the result of a competition between the rise of the entrance barrier, and the rise of the energy of the electrons trapped in the QD due to the capacitive cross-coupling between the entrance barrier gate and the QD. The dynamical nature of charge quantisation in the tunable-barrier pump was first recognized in an analysis of error mechanisms in the voltage-biased turnstile \cite{Zimmerman2004error}, and subsequently applied to the ratchet-mode pump\cite{fujiwara2008nanoampere,kashcheyevs2010universal}.

\subsection{The decay-cascade model}
We first described a scenario referred to as the “decay cascade model” \cite{fujiwara2008nanoampere,kashcheyevs2010universal, kashcheyevs2012quantum,kaestner2015non} developed in 2010--2015 to explain the trapping dynamics at the capture phase in a regime of strong coupling between the entrance barrier gate and the QD. We begin with an open QD, roughly as depicted in the first frame of figure \ref{Fig1} (a). The rising entrance barrier traps a number of electrons $n(t)$ in the QD in states separated by $E_{\text{C}}$, with the higher-$n$ states having a higher probability of tunneling back to the source. Note that in this formulation of the problem, the number of electrons in the QD can only decrease with time; tunneling into the QD from a thermal distribution in the leads is not considered. Mathematically, in the zero-temperature limit, we write the rate equation for the transition for $n$ to $n-1$ as $\text{d}P_{n} /\text{d}t= –\Gamma_{n} (t)P_{n}$ where $P_{n}$ is the probability that the QD contains $n$ electrons, $\Gamma_{n} (t)$ is a time-dependent escape rate after $t=t_{n}$ when the energy of the $n$-electron state rises above the Fermi level of the entrance lead. The trapping probability at a time $t$ is thus given by $P_{n} = \text{exp}[-\int_{t_{n}}^{t} \Gamma_{n}(\tau)\text{d}\tau]$.


\begin{figure}
\includegraphics[width=8.5cm]{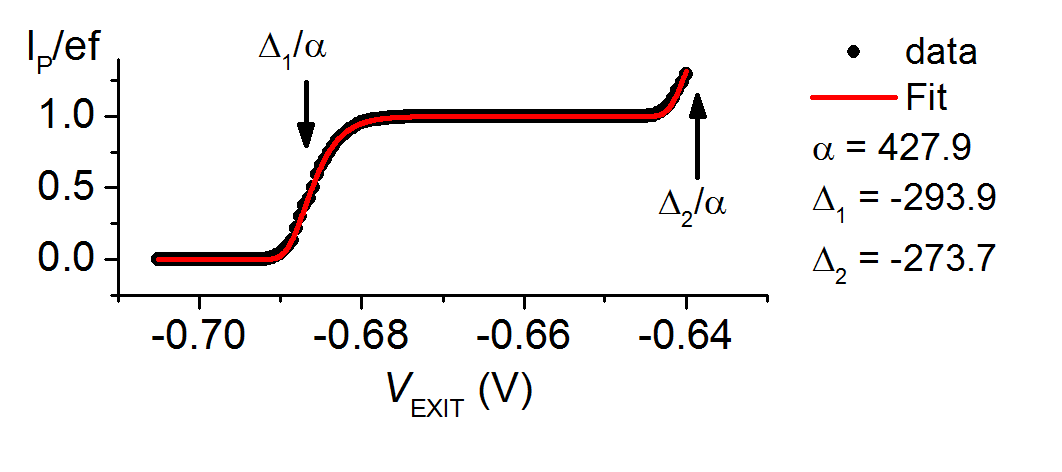}
\caption{\label{DecFit}\textsf{Example of fitting equation (1) to experimental data on a GaAs pump. The data is at $f=225$~MHz on a sample similar to the one described in Ref. \onlinecite{giblin2012towards}. Here, the exit gate voltage $V_{\text{EXIT}}$ takes the role of the tuning parameter $V_{\text{G}}$ in equation (1). The values of the 3 fit parameters are indicated on the plot. For this example, $\delta_{\text{2}} = \Delta_{\text{2}} - \Delta_{\text{1}}$ = 20.2}}
\end{figure}

The detailed form of $\Gamma_{n} (t)$ will depend on the shape of the entrance barrier, but most of the experimental results are consistent with the gate-defined tunnel barrier having a parabolic shape, which gives an exponential dependence of $\Gamma_{n}$ on the barrier height and time during the capture phase: $\Gamma_{n}(t) = \Gamma_{\text{a}} \text{exp}[-U_{n}(t)/k_{\text{B}}T_{\text{0}}]$, for $(T<T_{\text{0}})$, where $\Gamma_{\text{a}}$ is a constant, $U_{n}(t)=U_{\text{0}}(t) - nE_{\text{C}}$ is the linearly ramped barrier height as seen from the QD, and $T_{\text{0}}$  is an effective temperature characterising the energy dependence of tunneling probability, which corresponds to the crossover temperature of thermal hopping and tunneling\cite{yamahata2014accuracy}. Consequently, one can obtain an exponential solution for the integral part of $P_{n}$. For comparison with experimental data, a tuning voltage $V_{\text{G}}$ is introduced, which controls the QD potential. These assumptions lead to a generalised double-exponential formula for describing the pump current as a function of tuning voltage, up to arbitrary $n$ \cite{kashcheyevs2010universal}. Precision studies of the pump current have concentrated exclusively on the $N=1$ plateau, and the current as a function of tuning gate voltage has typically been fitted over a limited range of gate voltage to a simplified form of the full set of decay cascade equations presented in ref. \onlinecite{kashcheyevs2010universal}, for example\cite{giblin2012towards,giblin2017robust}:

\begin{eqnarray}
\frac{I_{\text{P}}}{ef} = \text{exp} \left[ -\text{exp} \left( -\alpha V_{\text{G}} + \Delta_{\text{1}} \right) \right] \nonumber \\ + \text{exp} \left[ -\text{exp} \left( -\alpha V_{\text{G}} + \Delta_{\text{2}} \right) \right],
\end{eqnarray}

where the three fitting parameters are $\Delta_{\text{1}}$, $\Delta_{\text{2}}$ and $\alpha$. Slightly different parameterisations of equation (1) were presented in refs. \onlinecite{bae2015precision} and \onlinecite{yamahata2016gigahertz}. The parameter $\delta_{\text{2}} = \Delta_{\text{2}} - \Delta_{\text{1}}$ (where $\delta_{\text{2}}$ has the same meaning here as in the full decay-cascade equations presented in ref. \onlinecite{kashcheyevs2010universal}) can be physically related to a ratio of tunnel rates\cite{kaestner2015non}, as discussed in more detail in sub-section D.

An example of a fit of equation (1) to some normalised $I_{\text{P}} (V_{\text{EXIT}})$ data is shown in figure \ref{DecFit}. Similar fits have played an important role in the development of tunable-barrier pumps in three main ways: Firstly, they provided support for the decay-cascade picture of QD initialisation. Equation (1), or equivalent parameterisations, yielded a good fit to $I_{\text{P}} (V_{\text{EXIT}})$ data on the $N=1$ plateau in several studies of ratchet-mode pumps \cite{giblin2010accurate,giblin2012towards,fletcher2012stabilization,bae2015precision,yamahata2016gigahertz,giblin2017robust}. Secondly, $\delta_{\text{2}}$ has been a commonly-used figure of merit for quantifying the flatness of the plateau as a function of QD depth-tuning parameter. A larger $\delta_{\text{2}}$ is equated with a flatter plateau\cite{kashcheyevs2010universal,giblin2012towards}. A single number, which can be extracted from a fit to easily acquired low-precision data is of great utility in evaluating a batch of pump samples, or in quantifying the change in plateau flatness as a function of magnetic field \cite{fletcher2012stabilization} or frequency \cite{giblin2012towards}. Thirdly, two studies\cite{giblin2012towards,yamahata2016gigahertz} have taken a further step, and used fits of equation (1) to estimate the optimal value of $V_{\text{EXIT}}$ for operating a pump, and the accuracy of the pump current at such an operation point. It was proposed that the inflection point of a fit to equation (1) could be defined as the optimal value of $V_{\text{EXIT}}$ for operating a pump\cite{giblin2012towards}. The deviation of equation (1) from $I_{\text{P}}/ef = 1$ at the inflection point is $\sim\!10^{-7}$ for $\delta_{\text{2}}=19$, and $\sim\!10^{-8}$ for $\delta_{\text{2}}=22$. For comparison, Ref. \onlinecite{giblin2012towards} reported $\delta_{\text{2}}=20$ at $f = 945$~MHz. Using a theoretical fit in this way to infer the accuracy of the pump at a certain operation point presupposes, of course, that the fit is based on a model which completely describes all the error processes. However, there is currently not sufficient data to justify this assumption. This is a key point, to which we will return in sections \ref{TuningSec} and \ref{ResultsSec}.



\subsection{Effect of magnetic field}
The precision measurements of GaAs pumps were enabled by the empirical discovery that the quantisation accuracy is dramatically improved by applying a magnetic field perpendicular to the plane of the sample\cite{wright2008enhanced,kaestner2009single}, and precision studies on GaAs pumps employed perpendicular fields of roughly $9$~T\cite{stein2016robustness}, $11$~T\cite{bae2015precision}, $13.5$~T\cite{giblin2017robust}, $14$~T\cite{giblin2012towards} and $16$~T\cite{stein2015validation}. The effect of magnetic field on Si pumps is less dramatic, and part-per-million accuracy has been achieved at $f=1$~GHz in zero field \cite{yamahata2016gigahertz,zhao2017thermal}. For GaAs pumps, the improvement in the quantisation with increasing field is generally considered to be due to the change in the tunnel rates caused by the additional confinement imposed on the electrons by the field\cite{fletcher2012stabilization}. Numerical calculations of the effect of the field on $\Gamma_{n}(t)$ were combined with the decay-cascade model to calculate $I_{\text{P}}(V_{\text{EXIT}})$ as a function of field, yielding results in rough qualitative agreement with experimental data which show, in one study, a roughly linear increase of the fit parameter $\delta_{\text{2}}$ with field\cite{fletcher2012stabilization,giblin2012towards}. 

\subsection{Extension of the decay-cascade model}
In this section, we discuss in more detail the meaning of the fit parameter $\delta_{n}$ in equation (1). This is an important exercise because, as already noted $\delta_{n}$ quantifies the plateau flatness, with a larger $\delta_{n}$ corresponding to a flatter plateau. The somewhat counter-intuitive relationship between the parameter labeled `$\delta_{\text{2}}$' and the flatness of the $N=1$ plateau is a consequence of terminology defined in Ref. \onlinecite{kashcheyevs2010universal}, which we follow here for consistency. There are two contributions to $\delta_{n}$\cite{kaestner2015non}:  

\begin{eqnarray}
\delta_{n} = \text{ln} \left[ \frac{\Gamma_{n}(t_{n})}  {\Gamma_{n-\text{1}}(t_{n})} \right] + \text{ln} \left[ \frac{\Gamma_{n-\text{1}}(t_{n})} {\Gamma_{n-\text{1}}(t_{n-\text{1}})} \right] \nonumber \\
= \text{ln} \left[ \frac{\Gamma_{n}(t_{n})}  {\Gamma_{n-\text{1}}(t_{n})} \right] + \text{ln} \left[ \frac{\Gamma_{n-\text{1}}(t_{n})}  {\Gamma_{n-\text{1}}(t_{n}+ \Delta t)} \right].
\end{eqnarray}

The first term of equation (2) is the simultaneous ratio $\Gamma_{n}(t) / \Gamma_{n-\text{1}}(t)$ governed by $E_{\text{C}}/k_{\text{B}}T_{\text{0}}$ and the second term is due to the decrease of $\Gamma_{n}$ during the time delay of the Fermi-level crossing of the QD levels, $\Delta t = t_{n-\text{1}} - t_{n}$. The second term depends on the magnitude of the cross-coupling as follows. We can express the rise speeds of the barrier potential and the QD as $\alpha_{\text{B}} \text{d}V_{\text{ENT}}/\text{d}t$ and $\alpha_{\text{I}} \text{d}V_{\text{ENT}}/\text{d}t$ respectively, where $\alpha_{\text{B}}$ and $\alpha_{\text{I}}$ are entrance-gate-voltage ($V_{\text{ENT}}$) to energy conversion factors. Thus, $\text{d}U_{n}(t)/\text{d}t = (\alpha_{\text{B}} - \alpha_{\text{I}})\text{d}V_{\text{ENT}}/\text{d}t$ and $\Delta t = E_{\text{C}} /(\alpha_{\text{I}} \text{d}V_{\text{ent}}/\text{d}t)$. Next, we introduce the cross-coupling parameter $g= \alpha_{\text{I}} / (\alpha_{\text{B}} - \alpha_{\text{I}})$ \cite{yamahata2014accuracy,oda2016nanoscale}, a device-dependent geometry parameter to characterize the competition between the rise of the QD and the barrier potentials. We can then write the increase of $U_{n} (t)$ during $\Delta t$ as $\Delta t \text{d}U_{n} (t) /\text{d}t =  E_{\text{C}} (\alpha_{\text{B}} -\alpha_{\text{I}})/\alpha_{\text{I}} = E_{\text{C}} /g$ and derive that $\Gamma_{n}$ decreases as $\text{exp} [-E_{\text{C}} /gk_{\text{B}}T_{\text{0}}]$, giving the second term in equation (2) as $E_{\text{C}} /gk_{\text{B}}T_{\text{0}}$. $g$ goes to infinity as $\alpha_{\text{I}}$ becomes close to $\alpha_{\text{B}}$, and $g$ becomes zero as $\alpha_{\text{I}}=0$. The more generalized energy-scale parameter $\Delta_{\text{PTB}}$ \cite{kashcheyevs2012quantum,kaestner2015non} (the rise of the QD potential during the decrease of escape rate by a factor of Euler’s number), is related to $g$ as $\Delta_{\text{PTB}} =gk_{\text{B}}T_{\text{0}}$. It defines a full decay cascade condition $\Delta_{\text{PTB}} \gg k_{\text{B}}T$ (equivalent to $g \gg T/T_{\text{0}}$) where the decay cascade model is valid in such a way that the electrons tunneling from the source to the QD do not play a significant role in the capture phase. In the tunnel decay cascade limit $(T< T_{\text{0}}, g \gg T/T_{\text{0}}), \delta_{n}$ is given by

\begin{equation}
\delta_{n} = \frac{E_{\text{C}}}{k_{\text{B}}T_{\text{0}}} + \frac{E_{\text{C}}}{\Delta_{\text{PTB}}} = \left( 1 + \frac{1}{g}\right) \frac{E_{\text{C}}}{k_{\text{B}}T_{\text{0}}}.
\end{equation}

Thus, the pump accuracy is independent of temperature, but it is determined by the charging energy, $T_{\text{0}}$, and the cross-coupling parameter. It might appear that a pump could be designed to operate in the decay cascade limit with arbitrarily high accuracy by making $g$ very small, but in practice a small $g$ implies that a large AC amplitude is required on the entrance gate to lift the electron over the exit barrier. The amplitude of $V_{\text{AC}}$ is limited in practice by heating\cite{chan2011single,yamahata2014accuracy}, which may violate the condition $\Delta_{\text{PTB}} \gg k_{\text{B}}T$, putting the pump in the thermal regime of operation discussed below. Rectification \cite{giblin2013rectification} may also impose practical limits to the amplitude of $V_{\text{AC}}$. It should also be noted that equation (3) has not been tested against experimental data in any of the high-precision studies. This would require measuring $E_{\text{C}}$ in the dynamical capture phase, as well as $T_{\text{0}}$. There have been a few reports of rough estimation of these parameters for silicon pumps: $E_{\text{C}}$ in the range 11-17 meV \cite{fujiwara2008nanoampere, yamahata2011accuracy, rossi2014accurate, zhao2017thermal} and $T_{\text{0}} \sim 20$~K \cite{yamahata2014accuracy}. 

\subsection{Thermal capture model}
Next, we discuss the second scenario of “thermal equilibrium capture” which describes the weak-cross-coupling limit $\Delta_{\text{PTB}} \ll k_{\text{B}}T$ (equivalent to $g \ll T/T_{\text{0}}$) where the electron-number states tend to be frozen to follow the initial grand canonical distribution of $P_{n}$ \cite{fricke2013counting}. In this limit, the transitions between current plateaus are symmetric as a function of the QD tuning gate, and the approximate expression when $E_{\text{C}}  \gg k_{\text{B}}T$ is given by the standard Coulomb blockade theory as

\begin{eqnarray}
\frac{I_{\text{P}}}{ef} = \sum_{n=1}^{\infty} n \left[ \frac{\text{exp}[-(E_{n} - n\alpha_{\text{G}}V_{\text{G}}) / k_{\text{B}}T]} {\sum_{n=1}^{\infty} \text{exp} [-(E_{n} - n\alpha_{\text{G}}V_{\text{G}}) / k_{\text{B}}T]} \right] \nonumber \\
\sim \sum_{n=1}^{\infty} \left[ 1 + \text{exp} \left(-\frac{E_{\text{C}}}{k_{\text{B}}T} \frac{V_{\text{G}}-V_{n}}{V_{n+\text{1}}-V_{n}}   \right) \right]^{-1},
\end{eqnarray}

where $\alpha_{\text{G}} = eC_{\text{G}} /C_{\Sigma}$, $E_{n} = n^{2}E_{\text{C}} /2$, and $V_{n+\text{1}} - V_{n} = E_{\text{C}} / \alpha_{\text{G}}= e/C_{\text{G}}$ as in equation (1). The pump accuracy is then governed by the conventional thermal-equilibrium parameter $E_{\text{C}}  / k_{\text{B}}T$. Thermal equilibrium electron capture has been observed in several experimental reports\cite{yamahata2014accuracy,zhao2017thermal} where the device features additional island gates which tend to reduce the cross-coupling. Note that more detailed theory\cite{fricke2013counting, kaestner2015non}, highlighting an increase of the effective energy separation between different electron number states, gives a correction to the Coulomb gap energy as $E_{\text{C}}  + [\Delta_{\text{PTB}} \text{ln} (\Gamma_{n+\text{1}} / \Gamma_{n})] =(1+g) E_{\text{C}}$. This is caused by the rise of QD potential during the characteristic time for freezing in each electron number state.

\begin{figure}
\includegraphics[width=8.5cm]{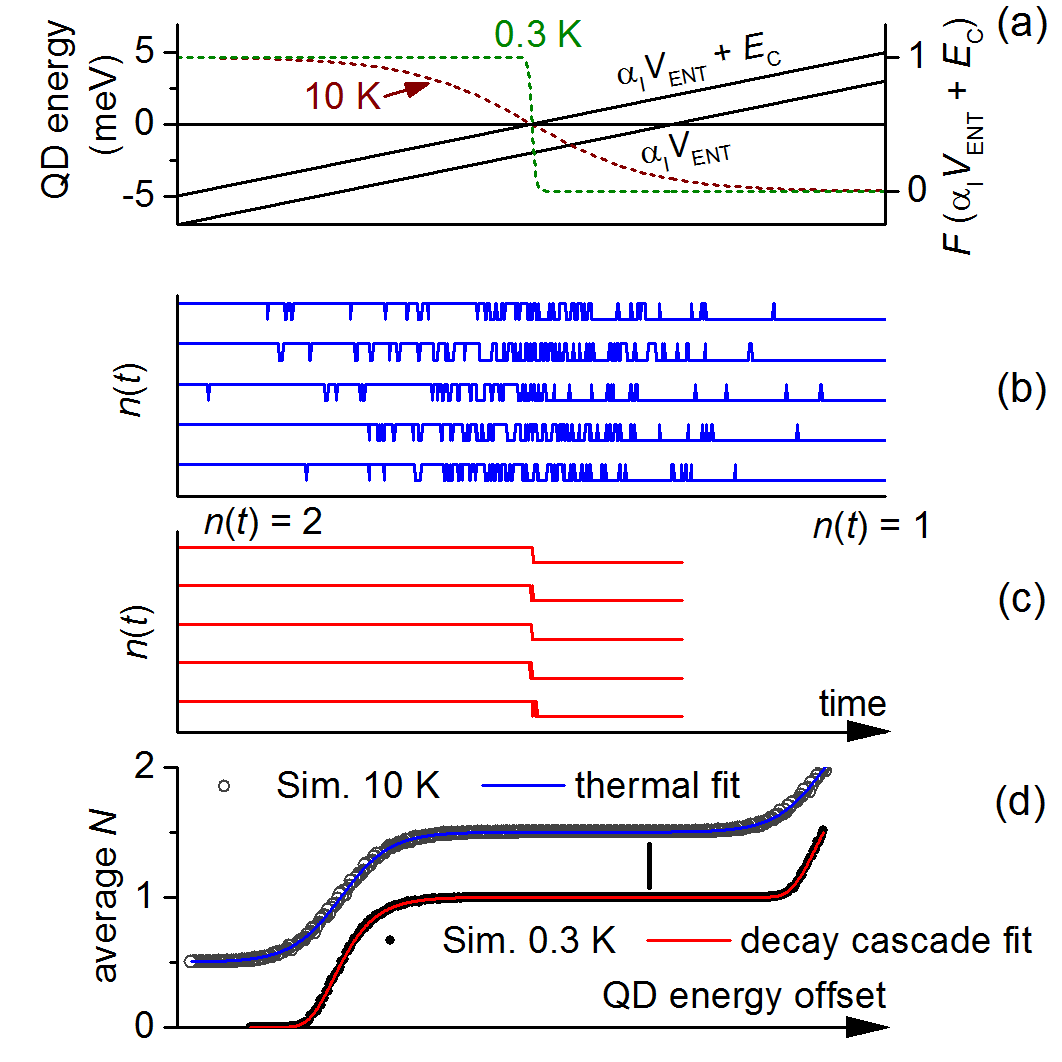}
\caption{\label{TheoryFig}\textsf{Panels (a)--(c) share the same x (time) axis. (a): Solid lines, left axis: time-dependence of the energy of the one- and two-electron states separated by $E_{\text{C}}=2$~meV used in a Monte-Carlo simulation of the initialization of the QD. A horizontal line indicates $E_{\text{F}}=0$. Dashed lines, right axis: Fermi functions at the time-dependent energy of the two-electron state for temperatures of $0.3$~K and $10$~K ($F=0.5$ when the energy of the two-electron state crosses the Fermi level) . (b): number of electrons in the QD as a function of time at the centre of the $N=1$ plateau at a temperature of $10$~K for 5 runs of the simulation. Traces are offset vertically for clarity. (c): As panel (b), but at a temperature of $300$~mK. (d): Simulated data for loading a QD at a temperature of $0.3$~K (filled points) along with a fit to equation (1), and $10$~K (open points, offset vertically by $0.5$ for clarity) with a fit to equation (4). Each data point is the average of $1000$ runs of the simulation. $\Delta_{\text{PTB}}=1$~meV for all simulations. }}
\end{figure}

\subsection{Monte Carlo simulation}
So far we have considered analytical solutions to rate equations, but insight into the QD loading process can also be gained from a simulation. In figure \ref{TheoryFig} we illustrate results from a simple Monte-Carlo simulation which calculates the probability of transitions between zero-, one- and two-electron states during the capture phase of the pump cycle, starting initially with a two-electron state. The QD energy increases linearly with time with respect to the Fermi energy of the lead, and the tunnel rates decrease exponentially with time. For each time step of the simulation, the probabilities of electrons tunneling into and out of the QD are calculated. Figures \ref{TheoryFig} (b) and (c) show the numbers of electrons in the QD, for five runs of the simulation, for a value of the exit gate (which tunes the QD potential) to load one electron, at temperatures of $10$~K and $0.3$~K, respectively. The other simulation parameters are $\Delta_{\text{PTB}} = 1$~meV, $E_{\text{C}}=2$~meV and $\Gamma_{\text{2}}(t)/\Gamma_{\text{1}}(t)=10^8$. From equations (2) and (3), these input parameters imply $T_{\text{0}}\sim\!1.4$~K. The two temperatures correspond roughly to the thermal, and decay cascade regimes respectively. The $n(t)$ simulations show a clear distinction between the two regimes: in the thermal regime, an electron is exchanged with the lead many times during the transition from the two- to the one-electron state, whereas in the decay cascade regime there is at most one into-dot tunneling event. The average number of electrons loaded into the QD, calculated from many runs of the simulation, is shown in figure \ref{TheoryFig} (d) as a function of QD depth-tuning parameter, along with fits to the first two summation terms of equation 1 (for the low-temperature decay cascade regime) and equation 4 (for the thermal regime). The fit to equation 1 yielded $\delta_{\text{2}}=20.3$, which is as expected from the simulation input parameters and the relation (see sub-section D above) $\delta_{\text{2}} = \text{ln}(\Gamma_{\text{2}}/\Gamma_{\text{1}}) + E_{\text{C}}/ \Delta_{\text{PTB}}$. The simulation shows clearly that the exchange of electrons with the lead in the thermal regime means that the Fermi distribution of electrons in the lead is reflected in the symmetric transition between plateaus, and helps to validate the assumptions underlying equations (1) and (4).

\subsection{Upper frequency limit}
As noted, equations 1 and 4 have successfully modeled the $I_{\text{P}} (V_{\text{EXIT}})$ dependence of a number of pumps, but they do not describe an intrinsic time-scale for the electron capture process, and so cannot model the frequency dependence of the pump accuracy. This is a problem of practical importance, as the accuracy of pumping has been observed to degrade with increasing frequency, limiting high precision studies to $f \lesssim 1$~GHz. Three possible mechanisms for frequency dependence have been proposed: source junction capacitance \cite{yamahata2015gigahertz}, dependence of the loading time on frequency\cite{ahn2017upper}, and nonadiabatic excitation \cite{kataoka2011tunable}. The first mechanism is due to the fact that electron capture is more likely to happen at earlier times in the pump cycle, corresponding to lower barrier heights. A dependence of barrier capacitance on barrier height could then result in a reduction in effective addition energy at higher pumping frequency\cite{yamahata2015gigahertz}. The second mechanism is motivated by the observation that increasing AC amplitude is required at the entrance gate to drive pumping as the frequency is increased\cite{ahn2017upper}. The third mechanism occurs when the shape of the QD potential is changing rapidly, and electrons enter a superposition of excited states which have a larger tunnel rate back to the source electrode. Clear signatures of this mechanism have been seen in some samples, where as $f$ is increased, the $N=1$ plateau is broken up into a number of sub-plateaus, each corresponding to an excited state \cite{kataoka2011tunable}. However, other nominally similar samples do not exhibit this behavior: The plateau continues to be described well by equation (1), with a $\delta$ fit parameter decreasing approximately linearly with frequency\cite{giblin2012towards}. A better understanding of excitation and relaxation processes in the dynamic QD is clearly needed, if the upper frequency limit of tunable-barrier pumps is to be extended further into the GHz range.

\subsection{Summary of theoretical understanding}
A reasonably robust theoretical picture of tunable-barrier pump operation has emerged. This theory considers the formation of a dynamic QD due to a rising entrance barrier, and yields the number of electrons remaining in the QD once the rising barrier has isolated it from the lead. The theory has received experimental support mainly by providing good fits to the pump current as a function of the QD depth-tuning gate (a role played by the exit gate in most experiments), both in the 'decay cascade' limit and the 'thermal equilibrium' limit. This type of data provides evidence that all the precision ratchet-mode tunable-barrier pumps studied to date are operating according to the same mechanism. However, an important question we should be able to ask is whether the theory can predict the accuracy of a given sample of a pump based on measurable parameters. Here, two important caveats must be considered. First, the parameters that enter the theory, tunnel rates and QD energies, are dynamic quantities which cannot be extracted from a series of DC characterization measurements. Physical parameters such as ratios of tunnel rates can be extracted from fits to experimental data, but the theory has not yet been able to make predictions analogous to the predictions possible for the adiabatic metallic pumps, described briefly at the beginning of this section. The second major caveat concerns the possible existence of error processes not described by the model. One such error process, non-adiabatic excitation \cite{kataoka2011tunable}, has been described above. Pumping through a parasitically formed QD\cite{rossi2018gigahertz} in addition to the `deliberate' QD could constitute another error process. The existence of these error process at an easily-measurable level in some samples raises the possibility that they may exist at some finite level in any sample. Thus, the theoretical models of QD initialisation, even though they may provide good fits to standard-accuracy characterisation data, cannot be assumed to be a reliable predictor of the pump accuracy at metrological levels. We return to this important point in section \ref{ResultsSec}.

\section{Characterisation of single-electron pumps\label{TuningSec}}

In this section, we review the methods used in Refs.~\cite{yamahata2016gigahertz,zhao2017thermal,giblin2012towards,bae2015precision,stein2015validation,stein2016robustness,giblin2017robust} to locate a region of externally adjustable parameters where precision measurements can be carried out. We assume from the outset that the gates have passed basic functionality tests, in other words that the source-drain conductance can be reduced to zero by adjusting $V_{\text{ENT}}$ and $V_{\text{EXIT}}$. In some device geometries it may be possible to obtain a set of $I-V$ curves as a function of a single gate voltage, preferably a plunger gate voltage which primarily has the effect of shifting the electrochemical potential of the QD but not substantially affecting the tunnel barriers. Such an experiment, typically known as source-drain bias spectroscopy, yields information about the charging energy of the dot, excited state energies, and the densities of states in the leads \cite{kouwenhoven1997electron}. However, this process did not form an important part of the tuning of the pumps used in the precision studies and in any case, two of the pump designs used for precision measurements (the NPL/Cambridge \cite{giblin2012towards} and PTB\cite{stein2015validation,stein2016robustness} designs) do not have separate plunger gates for addressing the QD energy independently of the barrier heights. We are not aware of any study in which parameters extracted from source-drain bias spectroscopy are correlated with high-accuracy pumping data. The pump operating point is determined empirically by measuring $I_{\text{P}}$ with the AC drive turned on.

\begin{figure}
\includegraphics[width=8.5cm]{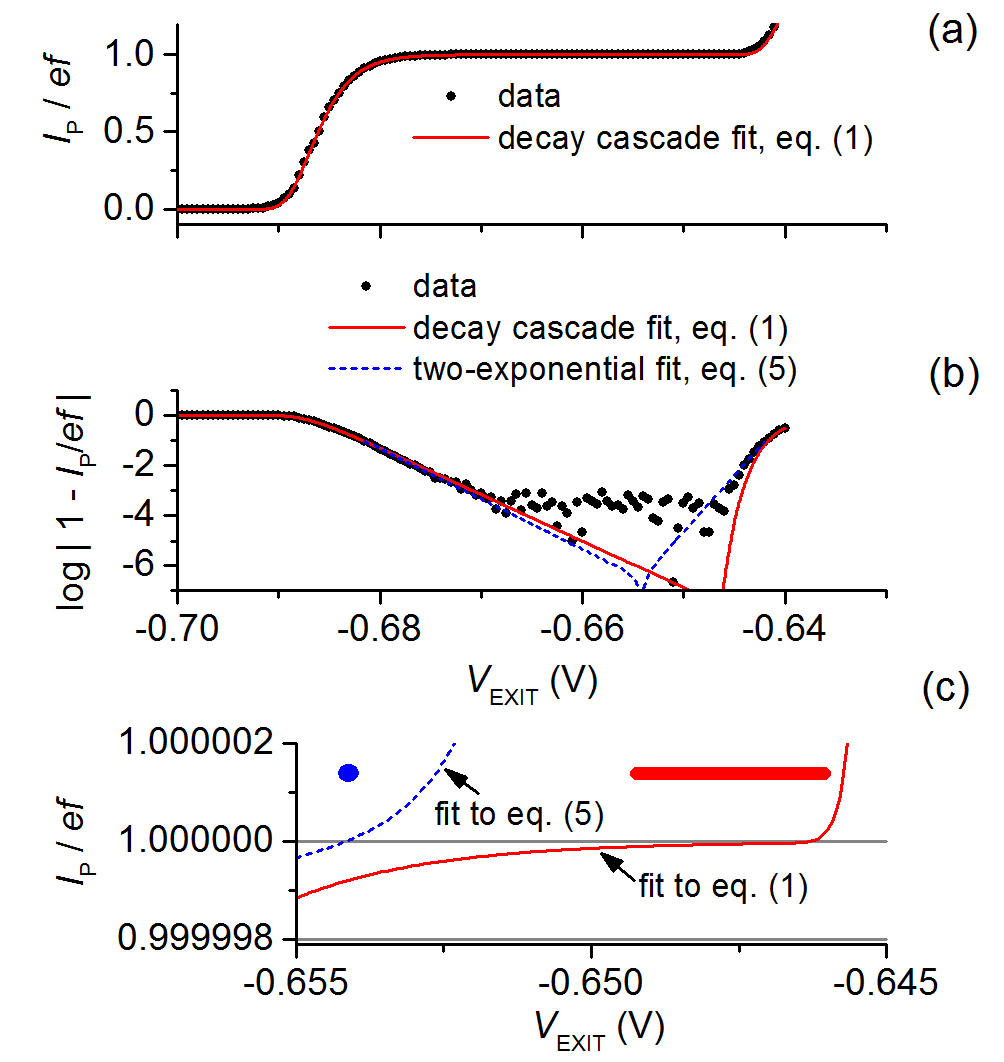}
\caption{\label{FittingFig}\textsf{(a): Normalised pump current (points) as a function of exit gate voltage for a GaAs pump similar to the one described in Ref. \onlinecite{giblin2012towards} at $f=225$~MHz, measured with a current preamplifier calibrated with an accuracy of $\sim 0.02 \%$. The line shows a fit to equation (1), with $\Delta_{\text{2}} - \Delta_{\text{1}} = \delta_{\text{2}}=20$. (b): The same data and fit as in panel (a), re-plotted to show deviations from perfect quantisation on a log scale. In addition, the blue dashed line shows a fit to equation \ref{eq:expfit}. (c): The fit lines from panels (a) and (b) plotted on an expanded linear scale. The filled point and thick horizontal bar show the values of $V_{\text{EXIT}}$ which would be selected for pump operation on the basis of two fits.}}
\end{figure}

The approximate operating point of the pump is found by measuring $I_{\text{P}}$ with modest ($\sim 0.1 \%$) accuracy as a function of $V_{\text{ENT}}$ and $V_{\text{EXIT}}$, over a fairly wide range of the gate voltages below the value at which the gate suppresses the channel conductance. This data set has become known as a `pump map' (figure \ref{Fig1}). An incomplete pump map often indicates an insufficiently large amplitude $V_{\text{AC}}$; increasing $V_{\text{AC}}$ extends the pumping plateaus along the $V_{\text{ENT}}$ direction\cite{kaestner2008robust}. For a two-gate pump, this process is relatively straightforward. As an example, the initial characterization measurements prior to the precision measurements reported in Ref. \onlinecite{giblin2012towards} with sine wave drive involved recording 5 pump maps with AC power levels of $-1, -1, -0.5, +1.2, +1.2$ dBm at the source, and frequencies of $150, 250, 400, 550, 630$~MHz. These measurements took a total of $\sim 3$~hours. Pumps with multiple gates to define and tune the QD \cite{giblin2017robust,zhao2017thermal} require a longer tuning process in which the extra gates are iteratively tuned to maximise the size of the $N=1$ plateau in the pump map. Optimisation of special drive waveforms using an AWG \cite{giblin2012towards,stein2015validation,stein2016robustness} also takes additional time, but again, the basic characterisation tool is the pump map. Additional insight into the plateau width can be gained by plotting the logarithm of the deviation of the current from $ef$. This type of plot was used as an aid to tuning a multi-gate pump \cite{giblin2017robust} and is also discussed in the next paragraph.

In several of the precision studies, the range of $V_{\text{EXIT}}$ for precision measurements and in some cases an optimal value of $V_{\text{EXIT}}$, was estimated from fits to $I_{\text{P}} (V_{\text{EXIT}})$ data. In refs. \onlinecite{giblin2012towards,bae2015precision,yamahata2016gigahertz}, the decay cascade model (see section \ref{TheorySec} B) was fitted to data over a wide range of $V_{\text{EXIT}}$ obtained with modestly-calibrated ($\sim 0.1 \%$) instrumentation. In ref. \onlinecite{zhao2017thermal} similar data was fitted to the thermal equilibrium capture model, a simplification of equation (4) limited to the $N=1$ plateau. In Refs. \onlinecite{stein2015validation,stein2016robustness}, data over a more limited range of $V_{\text{EXIT}}$ obtained with a precision measurement system was fitted to a phemonenological \textit{ansatz} consisting of a sum of two exponential functions\cite{kashcheyevs2014modeling}:

\begin{equation}\label{eq:expfit}
\frac{I_{\text{FIT}}(x)}{ef}=1+\delta_{\text{I}}-e^{-\alpha_1(x-x_1)}+e^{-\alpha_2(x-x_2)}.
\end{equation}

Here, $x$ is the scanned tuning parameter. This type of fit does not assume that the pump is operating in either the pure decay cascade or thermal equilibrium capture regimes. $\alpha_1$, $\alpha_2$, $x_1$, $x_2$, and $\delta_{\text{I}}$ are the fitting parameters. The parameter $\delta_{\text{I}}$ describes a possible offset of the plateau from $ef$: when fitting to this equation, we do not in general assume that the plateau is accurately quantised. The point of the minimum slope of the fit, also referred to as the point of inflection, was taken as the optimal value of $V_{\text{EXIT}}$ for the measurements with longest averaging time reported in ref. \onlinecite{stein2016robustness}. The location of the plateau may be defined as the range of the tuning parameter where $|1- I_{\text{fit}}/ef - \delta_{\text{I}} | < \delta_{\text{fit}}$, with $\delta_{\text{fit}}$ a few parts in $10^{8}$. This procedure was used in refs. \cite{stein2015validation, stein2016robustness,giblin2017robust}.

Fits of this type have utility in, for example, choosing a scan range for a high precision measurement of $I_{\text{P}} (V_{\text{EXIT}})$, but care should be taken not to assign any further significance to the fits. Different fits can give different estimates of the optimal operation point of the pump, as illustrated in figure \ref{FittingFig}. Here, an example $I_{\text{P}} (V_{\text{EXIT}})$ data set, obtained on a GaAs pump, has been fitted to equation (1) over the full range of plotted data, and to equation (5) over the range where $|1- I_{\text{P}}/ef| < 0.1$ (with $\delta_{\text{I}}$ constrained to zero). The ranges of $V_{\text{EXIT}}$ for which $|1- I_{\text{fit}}/ef | < 10^{-7}$ are indicated on figure \ref{FittingFig} (c). The fits clearly make very different predictions about both the flatness and the location on the $V_{\text{EXIT}}$ axis of the plateau. The log-scale plot, figure \ref{FittingFig} (b), highlights the fact that although equation (1) superficially fits the data well in figure \ref{FittingFig} (a), there are divergences between data and fit line on the approach to the plateau. Note that the fits assume a linear relationship between the scanned parameter and the QD energy, which does not generally hold to an arbitrary level of accuracy. 

\section{\label{MethSec}Precision measurement methods}

Here we address the metrological challenge posed by measuring currents of order $100$~pA with total relative uncertainties at or below the $10^{-6}$ level. Prior to the electron pump research effort, the lowest uncertainties available in sub nA current metrology were from reference current sources based on capacitor ramp techniques \cite{fletcher2007new,willenberg2003traceable,van2005accurate} which achieved uncertainties of a few parts in $10^{5}$. Early precision measurements of ratchet-mode pumps reached the limits of these reference sources \cite{giblin2010accurate}, and new strategies were needed.

\subsection{Measurement strategies}

The precision measurements of electron pump current considered in this paper have been traceable to the QHR and JVS, and therefore have some similarity with the classic `metrologial triangle' apparatus \cite{piquemal2000argument} which was proposed before the discovery of the ratchet-mode tunable-barrier pump. The studies reviewed in this article are not interpreted as `metrological triangle' experiments: they are simply measurements of an unknown current. In the apparatus described in Ref. \onlinecite{piquemal2000argument}, an electron pump current is compared with the QHR and JVS by using a cryogenic current comparator to amplify the pump current by a factor of $\sim\! 10^4$. The resulting microamp-level current is then passed through a QHR device and the Hall voltage measured in terms of the JVS. Several proof-of-concept experiments have been carried out in which CCCs were used to measure metallic \cite{steck2008characterization} and semiconducting \cite{kaestner2012characterization,giblin2016scaling} electron pumps, with relative uncertainties in the range $10^{-4}$ to $10^{-5}$. In the meantime, however, a fundamental problem with these CCC experiments came to light. Metrological CCCs have flux linkages $\sim\! 10$~$\mu$Aturns$/ \Phi_{0}$, where $\Phi_{0}$ is the magnetic flux quantum, so even with $50000$ turns (close to the practical limit), a current of $100$~pA generates a flux of only $0.5\times \Phi_{0}$. Evidence from back-to-back ratio accuracy tests (RATs) with low flux linkages suggests that rectification of noise by the CCC SQUID detector may generate errors at flux levels below $\sim\! 1 \mu \Phi_{0}$ \cite{drung2015improving}. At the very least, it would be difficult to convincingly verify that these errors were not present at the sub-ppm level in an electron pump measurement using a high-turns CCC. 

The alternative to directly scaling the pump current using a CCC, is to scale it indirectly using a resistor: the resistor is calibrated, typically using a CCC, at a high enough current such that the rectification of noise by the SQUID does not cause appreciable errors, and then used to measure the much smaller pump current under the assumption that the resistor does not have power- or voltage-dependence over the relevant range. This was the approach followed by the NPL and PTB groups, although the implementation differed. At NPL, the resistor was used in conjunction with a voltage source to generate a reference current equal in magnitude to the pump current (upper right inset in figure \ref{NoiseFig} (a), in which the electron pump is depicted as a current source). An ammeter calibrated to modest accuracy measured the small ($\lesssim 10$~fA) difference between pump and reference currents \cite{giblin2012towards}. In the case when the ammeter reads zero, the pump current is given by $I_{\text{P}} = V/R$. At PTB, the resistor was used as the feedback element in a trans-resistance amplifier (upper left inset in figure \ref{NoiseFig} (a)), with some important refinements to be detailed below. In this simple schematic circuit, we also have $I_{\text{P}} = V/R$ where $V$ is now the voltage at the amplifier output.

\subsection{Noise contributions}

\begin{figure}
\includegraphics[width=8.5cm]{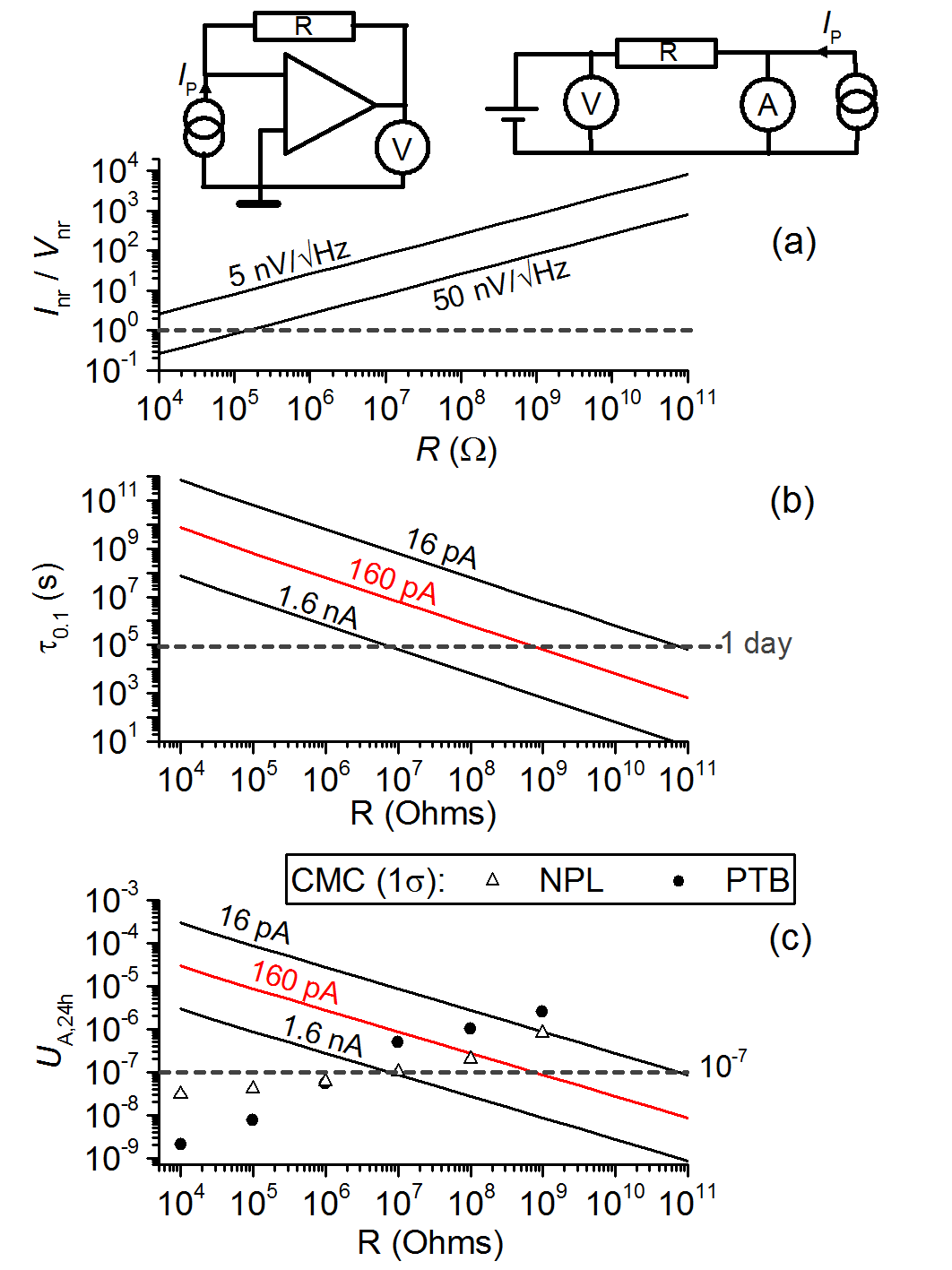}
\caption{\label{NoiseFig}\textsf{(a): Ratio of resistor thermal current noise to voltage noise as a function of resistance $R$, for two values of the voltage noise. The insets show schematic equivalent circuits illustrating the measurement of a current $I_{\text{P}}$ in terms of resistance and voltage. (b): time required to reach a type A uncertainty of $10^{-7}$ ($0.1$~ppm) for $3$ values of $I_{\text{P}}$, as a function of source resistance, assuming that the resistor thermal current noise is the only noise source. (c): lines: Type A uncertainty after $24$~hours as a function of source resistance. Open triangles and filled circles: calibration uncertainties for standard resistors at NPL and PTB respectively.}}
\end{figure}

Let us now discuss in some detail the noise considerations underlying the design of the NPL and PTB measurement systems. There are two significant sources of current noise in the circuits illustrated in the inset of figure \ref{NoiseFig} (a): in unit bandwidth, the thermal current noise in the resistor $\sqrt{4 k_{\text{B}}T/R}$ and the voltmeter noise $V_{n}/R$. The ammeter noise in the NPL setup, and the amplifier input noise in the PTB setup are much smaller and can be neglected. In unit averaging time, the thermal noise and voltmeter noise give respective contributions $I_{\text{nr}} = \frac{1}{I_{\text{P}}} \sqrt{4k_{\text{B}}T/R}$, and $V_{\text{nr}} = \frac{V_{n}}{I_{\text{P}}R}$ to the relative type A uncertainty of the measurement of $I_{\text{P}}$. 

Figure \ref{NoiseFig} (a) shows the ratio of these two noise contributions as a function of the resistor $R$ for two values of $V_{n}$, $50$~nV$/\sqrt{\text{Hz}}$ and $5$~nV$/\sqrt{\text{Hz}}$, the former being a worst-case value for a precision long-scale digital voltmeter (DVM), and the latter being a typical figure for a low-noise voltage pre-amplifier. It is clear that for $R \gtrsim 100$~k$\Omega$, the Johnson current noise is larger than the voltmeter noise, and in the following discussion we only consider $I_{n}$. Figure \ref{NoiseFig} (b) shows the averaging time required to reach a type A uncertainty of $10^{-7}$, $\tau_{\text{0.1}} = (I_{\text{nr}}/10^{-7})^2$, for a range of currents in the pA to nA range, and it is immediately clear that for the typical pump currents available from the present generation of pumps, the resistor needs to be around $1$~G$\Omega$ or more to avoid prohibitively long averaging times, and furthermore that averaging times of order 1 day will be required to reach metrological uncertainties. Figure \ref{NoiseFig} (c) shows the same information, here plotted as the type A uncertainty reached after $24$~hours of averaging, but also introducing an extra element to the problem: the type B uncertainty in calibrating the resistor. This is presented as the calibration and measurement capability (CMC) declarations of NPL and PTB. The measurement of $I_{\text{P}}$ is clearly a trade-off between using a lower-value resistor to minimize the type B uncertainty, and a higher-value resistor to minimize the type A uncertainty. 

The above discussion of averaging times has been simplified by assuming that the pump current is averaged continuously. This is not possible due to drifting offset currents and voltages in the measurement circuit, and measurements are performed using an on-off cycle which quadruples the time required to reach a given uncertainty for the simple reason that the pump current is only measured for half the time. The on-off cycle and the associated data analysis for extracting $I_{\text{P}}$ is described in the supplementary information to Ref. \onlinecite{giblin2017robust}, and further details of optimising the cycle, for example the use of auto zero in the voltmeter, is discussed in Ref. \onlinecite{stein2016robustness}. In precision electrical metrology, for example the calibration of standard resistors, it is usual to perform a measurement cycle in which the excitation current is reversed. Such a forward-reverse measurement cycle could in principle be implemented in the tunable-barrier pump, by exchanging the roles of entrance and exit gates. Because of the doubling of the size of the difference signal, the time taken to reach a given uncertainty would only be doubled, compared to a continuous measurement. This would introduce complications to the interpretation of the data because the two dynamic QDs formed by operating the device in two directions could be considered separate pumps with potentially different error processes, and for this reason a precision bi-directional pumping experiment has not yet been performed.

\subsection{Description of the NPL and PTB setups}

The measurement set-up used at NPL is schematically identical to the upper-right inset of figure \ref{NoiseFig} (a). It was based on pre-existing standards and calibration capability\cite{fletcher2000cryogenic}. A $1$~G$\Omega$ resistor was chosen for the first NPL measurement campaign\cite{giblin2012towards} which targeted a relative combined uncertainty of $10^{-6}$. Referring to figure \ref{NoiseFig} (c), a $100$~M$\Omega$ resistor provides a lower combined uncertainty than $1$~G$\Omega$ for $I_{\text{P}} \gtrsim150$~pA \cite{giblin2014sub}, but subsequent investigations showed that available thick-film $100$~M$\Omega$ standards did not have sufficient short-term stability to take advantage of the lower type B calibration uncertainty\cite{giblin2018limitations}, and the $1$~G$\Omega$ was retained for subsequent measurement campaigns. A precision DVM (HP 3458A) fulfilled the function of the voltmeter. Initially \cite{giblin2012towards} the DVM was calibrated at $1$~V using a transportable DC reference voltage, but for later measurement campaigns \cite{bae2015precision,yamahata2016gigahertz,giblin2017robust,zhao2017thermal} it was calibrated directly against a primary Josephson voltage standard (JVS). A computer controlled low-thermal switch (Data Proof DP320) allowed connection of the DVM to the JVS without powering down or moving the DVM, and daily calibrations reduced the contribution of the voltage measurement to the overall relative uncertainty to $10^{-7}$ or less. For most of the measurement campaigns\cite{giblin2012towards,bae2015precision,yamahata2016gigahertz,giblin2017robust}, the largest contribution to the uncertainty budget was the $8 \times 10^{-7}$ relative CMC uncertainty in calibrating the $1$~G$\Omega$ resistor. A comprehensive re-evaluation of this uncertainty \cite{giblin2018reevaluation} resulted in a lower value, $\sim 10^{-7}$ and this was applied in the most recent measurement campaign, resulting in the lowest relative combined uncertainty for the NPL system \cite{zhao2017thermal} of $2.7 \times 10^{-7}$.

While the NPL measurement system was constructed around existing standards and instruments, the PTB measurement system used for the measurements of Refs. \onlinecite{stein2015validation,stein2016robustness} used a new, specially designed instrument, the ultrastable low-noise current amplifier (ULCA) \cite{drung2015ultrastable}. Overall, the ULCA functions as a transresistance amplifier, as shown in the upper left inset of figure \ref{NoiseFig} (a), with a nominal gain of $10^{9}$ V/A. However, internally it contains two functional blocks: an input current gain stage with a gain of $1000$ and a nominal input resistance of $3$~G$\Omega$, followed by a transresistance stage with a $1$~M$\Omega$ feedback resistor. Referring to figure \ref{NoiseFig} (c), this two-stage design allows the trans-resistance gain to be traceably calibrated against the resistance scale with the low uncertainty of a $1$~M$\Omega$ resistor, while maintaining the favorably low input input noise of the $3$~G$\Omega$ input stage. The overall trans-resistance gain of the ULCA can be calibrated using a CCC, in two steps\cite{drung2015improving,stein2016robustness,giblin2019interlaboratory}, with a relative combined uncertainty $\sim\!2 \times 10^{-8}$.

\begin{table}
\caption{\label{NPLtable}Uncertainty budget for the NPL measurement system, in parts in $10^{6}$ taken from Ref. \cite{zhao2017thermal}. Only the 4 largest uncertainty contributions are shown. $I_{\text{P}} \sim 160$~pA.}
\centering
\setlength{\tabcolsep}{8pt}
\begin{tabular}{c c c}
\hline\hline
Type A (10 hours averaging) & 0.229 \\ 
Uncertainty of 1 G$\Omega$ calibration & 0.1 \\ 
1  G $\Omega$ short-term drift & 0.07 \\ 
DVM drift between calibrations & 0.068 \\ 
\hline
Total of these 4 terms & 0.268 \\
Total published uncertainty & 0.27
\end{tabular}
\end{table}

\begin{table}
\caption{\label{PTBtable}Uncertainty budget for the PTB measurement system, in parts in $10^{6}$ taken from Ref. \cite{stein2016robustness}. Only the 4 largest uncertainty contributions are shown. $I_{\text{P}} \sim 96$~pA.}
\centering
\setlength{\tabcolsep}{8pt}
\begin{tabular}{c c c}
\hline\hline
Type A (21 hours averaging) & 0.13 \\ 
Stability of ULCA between calibrations & 0.08 \\ 
Uncertainty of ULCA calibration & 0.015 \\ 
Miscellaneous ULCA effects & 0.014 \\ 
\hline
Total of these 4 terms & 0.154 \\
Total published uncertainty & 0.16
\end{tabular}
\end{table}

In tables \ref{NPLtable} and \ref{PTBtable}, we list the four largest uncertainty contributions to the measurement of $I_{\text{P}}$, in the lowest-uncertainty measurements reported by, respectively, the NPL and PTB groups. The type A uncertainty is the largest contribution for both the measurement systems. For the NPL system, this is dominated by the thermal noise in the $1$~G$\Omega$ reference resistor. The ULCA used in the PTB measurement has a factor $\sqrt{3}$ lower input noise, and the PTB system gains an additional factor $\sqrt{2}$ by virtue of using two ULCAs, one on each side of the pump. Thus, for a given averaging time, the PTB system has a factor $\sim\!2.4$ times lower type A uncertainty than the NPL system. The absence of any significant contributions in the PTB table due to the voltage measurement is because the output voltage of the ULCA is opposed by the voltage from a JVS, with a voltmeter recording the small residual difference signal. It is noteworthy that the stability of instruments in between calibrations is a significant contributor to both uncertainty budgets: this shows the extent to which the electron pump measurements have pushed the limits of electrical metrology, with the pumps themselves being arguably the most stable standards in the experiments. In the NPL system, the stability of the $1$~G$\Omega$ resistor is a limiting factor \cite{giblin2018limitations,giblin2019interlaboratory}, and it is unlikely that the overall uncertainty can be pushed much below $2 \times 10^{-7}$. The prospect of the PTB system yielding significantly lower uncertainties than the benchmark $1.6 \times 10^{-7}$ depends on reduction of the type A uncertainty, and improvement of the stability of the gain in between calibrations. Both are active development areas, especially the former, which is addressed using specialized ULCAs having larger input resistances than the standard $3$~G$\Omega$ \cite{drung2017ultrastable}.

\begin{table*}[]
\centering
\begin{tabular}{|c| |c| |c| |c| |c| |c| |c| |c|} 
 \hline
 ~ & Ref.~\cite{giblin2012towards}  & Ref.~\cite{bae2015precision}  & Ref.~\cite{giblin2017robust}  & Ref.~\cite{stein2015validation} & Ref.~\cite{stein2016robustness} & Ref.~\cite{zhao2017thermal} & Ref.~\cite{yamahata2016gigahertz}   \\ [0.5ex] 
 \hline\hline
 Gate Voltages & 1 of 2 & 1 of 7 & 3 of 7 & 1 of 2 & 2 of 2 & 2 of 7 & 1 of 3 \\ 
 $B$-field & $X$ & $X$ & $X$ & $X$ & 12, 14, 16 T & $X$ & $X$ \\
 $P_\textup{AC}$ & $X$ & $X$ & 3--6 dBm & $X$ & $X$ & $X$ & $X$\\
 $T_\textup{bath}$ & $X$ & $X$ & $X$ & $X$ & $X$ & $X$ & $X$\\
 $f$ & 400, 630, 945 MHz & $X$ & $X$ & $X$ & 100--600 MHz & $X$ & 1, 2 GHz \\
 Drain/Source Bias & $X$ & $X$ & $X$ & $X$ & -10--10 mV & $X$ & $X$\\ [1ex] 
 \hline
\end{tabular}
\caption{Robustness table. Experimental parameters that have been varied to study the robustness of the current quantization in high-accuracy experiments. An $X$ indicates that the parameter has been kept at a single fixed value.}
\label{RobustTable}
\end{table*}

\begin{table*}[]
\centering
\begin{tabular}{|c| |c| |c| |c| |c| |c| |c| |c|} 
 \hline
 Material & $f$~(MHz) & $\Delta I_\textup{P}\pm u_\textup{T}$~(ppm) & Measurement/ & Waveform & plateau average (PA) & T(K) & B(T) \\ [0.5ex] 
~ & ~ & ~ & Fab. Lab. & ~ & or fixed point (FP) & ~ \\ [0.5ex]
 \hline\hline
 Si\cite{zhao2017thermal} & 1000 & $-0.26\pm 0.27$ & NPL/UNSW & Sine & PA & $0.3$ & $0$ \\ 
 Si\cite{yamahata2016gigahertz} & 1000 & $-0.64\pm 0.92$ & NPL/NTT & Sine & PA & $\sim 1.5$ & $0$ \\
 GaAs\cite{bae2015precision} & 950 & $-0.92\pm 1.37$ & NPL/KRISS & Sine & PA & $0.3$ & $11$ \\
 GaAs\cite{giblin2012towards} & 945 & $-0.51\pm 1.20$ & NPL/Cambridge & AWG & PA & $0.3$ & $14$ \\
 GaAs\cite{stein2016robustness} & 600 & $-0.10\pm 0.16$ & PTB/PTB & AWG & FP & $0.1$ & $\sim 9$ \\
 GaAs\cite{stein2015validation} & 545 & $-0.06\pm 0.20$ & PTB/PTB & AWG & PA &$ 0.1$ & $16$ \\
 GaAs\cite{giblin2017robust} & 500 & $+0.28\pm 0.86$ & NPL/KRISS & Sine & FP & $1.3$ & $13.5$ \\ [1ex] 
 \hline
\end{tabular}
\caption{Agreement table. AWG = arbitrary waveform generator.}
\label{UniTable}
\end{table*}

\section{\label{ResultsSec} Results of precision measurements}

In this section, we discuss the precision measurements, and the evidence for agreement and robustness. The precise meaning of these terms was stated in the introduction: `agreement' means that all pumps generate the same current within the measurement uncertainty, and `robustness' means that any one pump generates a constant current even if its control parameters are varied. 

\subsection{Evidence for robustness}

We turn our attention to robustness first, because any statement about agreement at a given uncertainty level presupposes that the devices involved in the experiments have already demonstrated robustness, i.e, invariance of the current as a function of tuning parameters, at that uncertainty level. Table \ref{RobustTable} summarises which tuning parameters in each of the seven studies were adjusted to investigate the invariance of the pump current against that parameter. The only parameter which was systematically investigated in all the studies is $V_{\text{EXIT}}$, which, as we have seen in section \ref{TheorySec}, is a key parameter for interpreting the capture mechanism. The reason for the sparse population of the table is that, as is clear from section \ref{MethSec}, measuring one data point with a relative type A uncertainty of $10^{-7}$ can take a time of order 1 day. A thorough investigation of the robustness of a pump against all of its tuning parameters is therefore a considerable undertaking. The simplest possible design of two-gate pump has, in addition to $V_{\text{ENT}}$ and $V_{\text{EXIT}}$, the amplitude of $V_{\text{AC}}$ as a third tuning parameter. The magnetic field, if applied, is a fourth parameter, and the source-drain bias, although nominally zero, should also be considered, making a minimum set of $5$ parameters. One of the silicon pump designs\cite{yamahata2016gigahertz} has a third gate on top of the device to induce carriers into the device channel, and two of the pump designs\cite{bae2015precision,giblin2017robust,zhao2017thermal} have a number of additional gates to confine the electrons in the QD and provide additional fine-control of the pump tuning. There is clearly a trade-off between simplicity and tunability, given that a convincing demonstration of robustness should include all control parameters.

Two of the studies \cite{stein2016robustness,giblin2017robust} reported robust plateaus as a function of several control parameters. In ref. \onlinecite{giblin2017robust}, plateaus were reported as a function of the voltage at three control gates, and the amplitude of the AC entrance gate drive, with relative uncertainties of $\sim\!2 \times 10^{-6}$. Reference \onlinecite{stein2016robustness}, the lowest-uncertainty robustness study to date, reported plateaus in $V_{\text{ENT}}$ and $V_{\text{EXIT}}$ with relative uncertainties of $\sim\!6 \times 10^{-7}$, a plateau in magnetic field with relative uncertainty $\sim\!4 \times 10^{-7}$ and a plateau in source-drain bias voltage with relative uncertainty $\sim\!2.5 \times 10^{-7}$. Here, we refer to the uncertainty for each data point. Robustness cannot be inferred at a lower uncertainty than the type A uncertainty for each data point in the plateau scan\footnote{In measuring robustness, we do not need to know the absolute value of the current so the type B uncertainty is not important. This type of study does not require an accurately calibrated measurement system, as long as it is stable}. This is a very important point which must be stressed: a robustness study, by definition, is an empirical exercise which should not assume \textit{a priori} any functional form to the data, either based on theory or an empirical ansatz. To emphasize this point by way of an example, we can imagine a scenario where measurements of $I_{\text{P}} (V_{\text{EXIT}})$ with relative uncertainty of $10^{-6}$ yielded a good fit to equation (5), but the pump had a $V_{\text{EXIT}}$-dependent error of a few parts in $10^{7}$ due to rare pumping through a second parasitic QD. The errors would not be resolved by the measurements, and the experimenter might conclude erroneously by following the fit line that the pump was accurate at the $10^{-7}$ level in the middle of the plateau.

As already noted, the measurement time required for a comprehensive robustness study is considerable: each data point requires roughly 1 day of integration to reach a relative uncertainty of $10^{-7}$. To authors are not aware of any such study being published to date. The general verification of robustness at the this level is a minimal requirement for the widespread operation of electron pumps as primary current standards, and this remains a subject for future work. The limitations of the available robustness data will be an important consideration below, where we consider the agreement between different electron pump designs.

\subsection{Evidence for agreement}

\begin{figure}
\includegraphics[width=8.5cm]{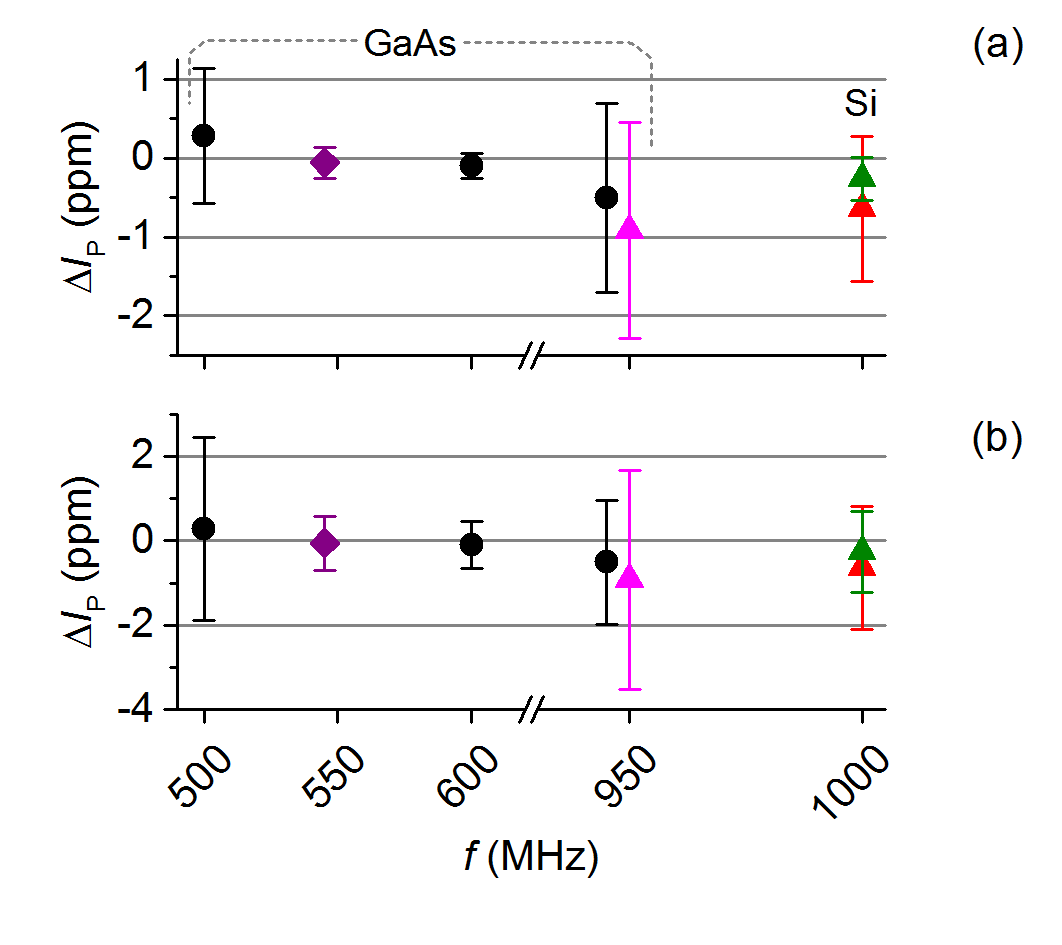}
\caption{\label{UnivPlot}\textsf{(a): Normalised deviation of pump current from $ef$, as reported in the seven studies considered in this review: $\Delta I_{\text{P}} = (I_{\text{P}} - ef) / ef$. Error bars are the $1\sigma$ total uncertainty, as reported. (b): the same data from plot (a), with error bars expanded to include the average type A uncertainty of the single data points plotted in figure \ref{ExitGateFig}. Note the different y-axis scales of plots (a) and (b).}}
\end{figure}

The seven high-precision studies have all presented top-level results in the form of a single number: the deviation of the pump current from its expected value of $ef$, either with the pump in a single optimally tuned state or averaged over a range of states with a tuning parameter varied. These important results are tabulated in table \ref{UniTable}, in decreasing order of pump operating frequency, and plotted in figure \ref{UnivPlot} (a). It is clear from this summarised data that the seven studies, on 5 different designs of pump made in different fabrication laboratories, have all reported currents equal to $ef$ within relative uncertainties at or below $10^{-6}$. From this observation, we could draw the encouraging conclusion that devices implementing the ratchet mode of pumping are capable of metrological accuracy, regardless of the details of device design and fabrication. However, to interpret this data we need to consider first the detailed differences in the analysis methods used to arrive at the data points of figure \ref{UnivPlot}(a), and secondly, the extent to which the pump in each study was shown to exhibit robustness.

All of the seven precision studies presented a high-resolution plot demonstrating a plateau in $I_{\text{P}} (V_{\text{EXIT}})$. These data are compiled in figure \ref{ExitGateFig}. In five of the studies, the single value of the pump current shown in figure \ref{UnivPlot}(a) was averaged over several measurements at different values of $V_{\text{EXIT}}$. The type A uncertainty in the averaged value, evaluated as the standard error on the mean, was smaller than the uncertainty on each data point by a factor 2-3. The method of choosing the range of $V_{\text{EXIT}}$ constituting the plateau differed between the studies. In the first three NPL measurements \cite{giblin2012towards,bae2015precision,yamahata2016gigahertz}, following an earlier, lower-precision study\cite{giblin2010accurate}, a statistical method was used in which the plateau was defined as a range of data points in the $I_{\text{P}} (V_{\text{EXIT}})$ plot for which the gradient of a linear fit is zero within the uncertainty of the fit. This method still allows some subjective judgment as to which set of points to choose, as more than one range of points will satisfy the linear fit criterion. The other method which has been used to select a plateau is based on fits to the normalised current $I_{\text{P}}/ef$ over a wider range of $V_{\text{EXIT}}$. Here, as discussed in section \ref{TuningSec}, the plateau is defined as the range of $V_{\text{EXIT}}$ over which the fit line deviates from $I_{\text{P}}/ef = 1$ by less than some small amount $\delta_{\text{fit}}$. This method was used in Ref. \onlinecite{zhao2017thermal} using a fit to equation (4), with $\delta_{\text{fit}} = 3 \times 10^{-8}$, and in Ref. {\onlinecite{stein2015validation} using a fit to equation (5) with $\delta_{\text{fit}} = 1 \times 10^{-8}$. The two methods for selecting a plateau (based on linear fits and exponential model fits) were compared in Ref. \onlinecite{giblin2017robust}, and it was shown that they generally give the same average current, at the $10^{-6}$ level. 

In contrast to averaging over a plateau, two of the studies \cite{stein2016robustness,giblin2017robust}, presented $I_{\text{P}}$ averaged for long periods (48 and 21 hours respectively) at a fixed pump operating point. This optimal point on the $V_{\text{EXIT}}$ axis was chosen in Ref. \onlinecite{stein2016robustness} as the point of inflection of the two-exponential fit (equation (5)) to a range of data. In Ref. \onlinecite{giblin2017robust} the choice of operating point was slightly more subjective, based on plotting the deviation of the current from $ef$ on a log scale, and iteratively tuning the multiple gates of the device.


\begin{figure}
\includegraphics[width=8.5cm]{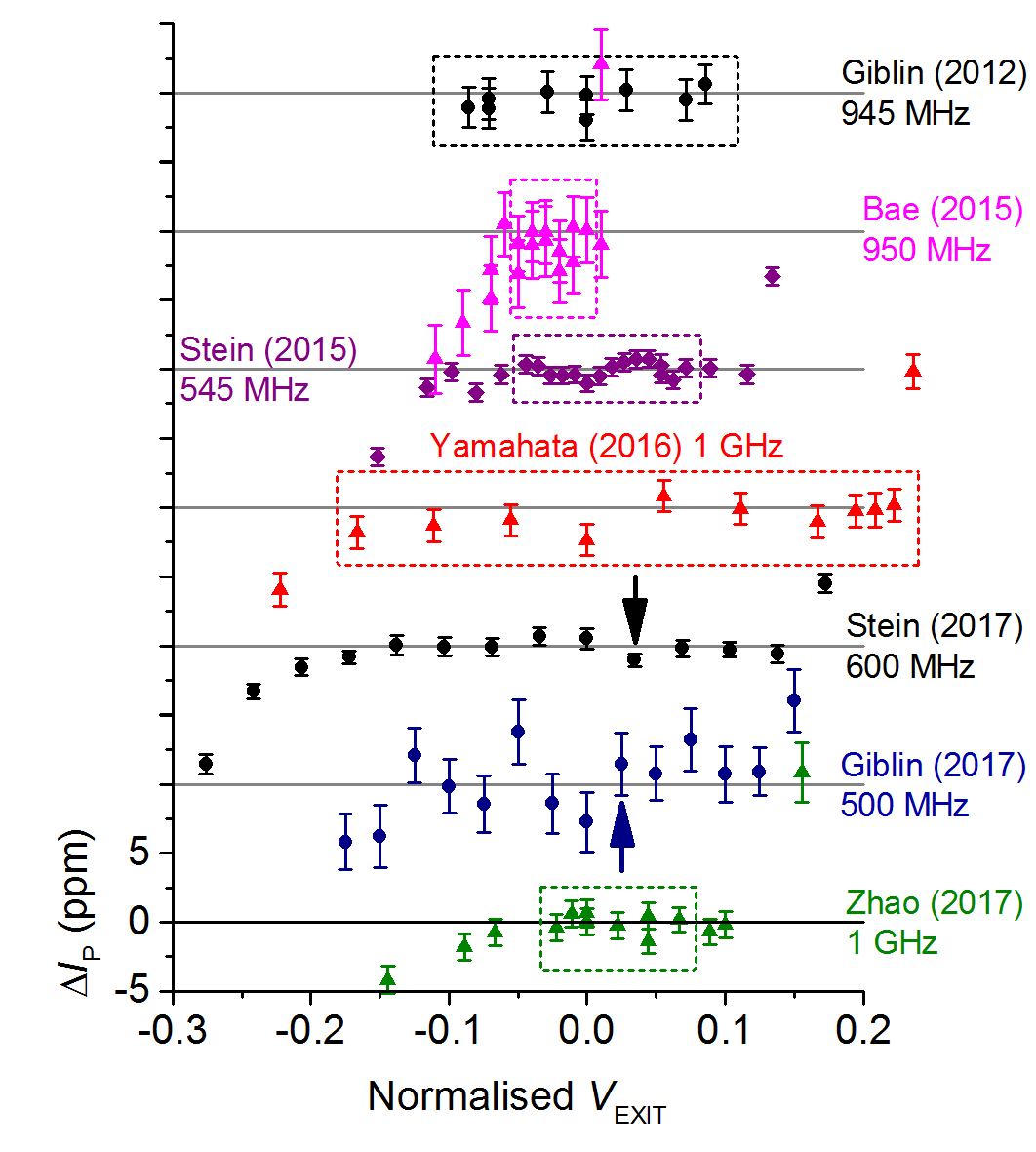}
\caption{\label{ExitGateFig}\textsf{Normalised deviation of pump current from $ef$, $\Delta I_{\text{P}} = (I_{\text{P}}-ef)/ef$ plotted against normalised exit gate voltage, for each of the high-precision studies in this review. The exit gate voltage is normalised to the span of exit gate for which $|I_{\text{P}}-ef|/ef<0.1$, and the plots are offset vertically by multiples of $10$~ppm for clarity. Dashed boxes indicate the data points averaged to yield the single data point in figure \ref{UnivPlot} for the data of Refs. \onlinecite{giblin2012towards,bae2015precision,stein2015validation,yamahata2016gigahertz,zhao2017thermal}, and short arrows indicate the value of $V_{\text{EXIT}}$ for the long measurement yielding the data points in figure \ref{UnivPlot} for Refs. \onlinecite{stein2016robustness,giblin2017robust}. The colour coding of the data points matches figure \ref{UnivPlot}}}
\end{figure}

Irrespective of the method used to determine the operating point, or range of points, it follows from the above section that any statement we can make about the agreement of the different pumps is currently limited by the uncertainties in the individual measurements (the data of figure \ref{ExitGateFig}) used to determine the plateau robustness. With this in mind, the averaged `headline' data points of figure \ref{UnivPlot}(a) have been re-plotted in figure \ref{UnivPlot}(b) with the error bars expanded to include the average type A uncertainty of the individual data points in the $I_{\text{P}} (V_{\text{EXIT}})$ scans of figure  \ref{ExitGateFig}. In other words, these are the uncertainties justified by a strictly empirical interpretation of the robustness (as a function of $V_{\text{EXIT}}$ only) data. Of course, for most of these studies, the robustness as a function of other parameters was not verified, but nevertheless figure \ref{UnivPlot} (b) represents a reasonable unbiased comparison of the available precision data.


In the context of the agreement between different pumps, it is also important to note that the authors are not aware of blind-test methodology being used in any of these studies. In other words, the experimenters performing precision pump measurements knew the results of their measurements (deviation of pump current from $ef$) during the measurement campaigns. The metrology community is generally aware of the problem posed by experimenter bias, and at least one recent precision measurement of the Planck constant \cite{schlamminger2015summary} employed a blind test methodology to eliminate bias. In the electron pump measurement, bias could enter through the settings of gate voltages or AC amplitude which are kept fixed during the experiments, possibly in a passive way (if the current is found equal to $ef$ in the first experiment, no attempt is made to adjust the parameters). This consideration highlights the case already made, for efforts to be devoted to a thorough investigations of robustness in all control parameters.

\subsection{Discussion and future work}

We have reviewed in detail the published precision data obtained in seven experiments on five different types of tunable-barrier pumps, carefully tuned by a variety of methods. As presented, the lowest-uncertainty measurements indicate an agreement between a GaAs pump \cite{stein2016robustness} and a Si pump \cite{zhao2017thermal} within a combined uncertainty of $0.3$~ppm. These experiments achieved their low uncertainties through a combination of low type-B uncertainties, and statistical averaging at a fixed operating point\cite{stein2016robustness}, or over a range of operating points \cite{zhao2017thermal}, to reduce the type A uncertainty. The choice of operating point(s) on the $V_{\text{EXIT}}$ axis was in turn justified by extrapolation of fits to $I_{\text{P}} (V_{\text{EXIT}})$ data sets. If we treat the high-precision data as a purely empirical study of the pump accuracy as a function of $V_{\text{EXIT}}$, the uncertainty of each data point in the scan limits the total uncertainty, and the combined uncertainty in the agreement between Refs. \onlinecite{stein2016robustness,zhao2017thermal} increases to about $1$~ppm as shown in figure \ref{UnivPlot} (b).

To return to the question posed at the start of this review, the available data on agreement and robustness constitute promising evidence that the tunable-barrier pumping mechanism can potentially be implemented in a universal way, at the $1$~ppm uncertainty level. A related, and more difficult, question, is to what extent the pumps can already be considered primary current standards. In posing this question, we are imagining a future scenario in which a laboratory which does not already have access to accurate electric current traceability could take a sample of electron pump, perform an agreed characterization and tuning procedure, and then treat the device as a primary standard. This scenario would constitute using the electron pump as a primary standard in an analogous manner to the present use of the QHE and JVS. A key ingredient of this scenario is a characterization procedure, and before this can be agreed upon, the high-precision data set on electron pumps needs to be expanded in several dimensions: more rigorous explorations of the robustness of pumps as a function of all the available tuning parameters, investigations of multiple samples of pumps, on multiple cool-downs and critically, demonstration of a reliable extrapolation between the wide-range features of the data accessible with standard instrumentation and the accuracy of the pump at metrological uncertainty levels. A large enough data set, showing correlation between high-accuracy operation and certain features of characterization data (such as flat plateaus measured with low precision), may enable at least some designs of pumps to be considered suitable for use as current standards. The accumulation of large precision data sets is a formidable challenge in terms of measurement time, but on the other hand, the infrastructure required for these measurements (a $1.5$-K cryostat without a magnet can be sufficient\cite{yamahata2016gigahertz}) is modest compared to some metrological undertakings, making it possible for a broad range of laboratories to participate in this research. Alternatively, the self-referenced current standard \cite{fricke2013counting,fricke2014self} presents a powerful method for utilising an electron pump without reference to a characterization procedure. With this method, electron transport errors are counted in real time, so the demands on pump transfer accuracy are relaxed. As noted in the introduction, however, making this method work at metrologically-useful current levels presents a considerable technical challenge.

\section{\label{ConcSec} Conclusions}

The success of the tunable-barrier pump in generating an accurate quantised current would have been difficult to predict based on the first measurements on semiconductor pumps in the early 1990s, and is a tribute to major efforts in fabrication, small current metrology and theoretical understanding. Uncertainties in current measurements have reduced from $10^{-4}$ in 2008, to $10^{-7}$ at the present time, and at each stage the tunable-barrier pumps have been found to generate currents equal to $ef$ within the measurement uncertainty. The accumulated precision electron pump measurements are certainly encouraging from the perspective of future development of the electron pumps as current standards. Seven studies have shown plateaus in at least one control parameter, the exit gate, where the current is equal to the expected value $ef$ within a relative combined uncertainty of around $10^{-6}$. Two studies have shown plateaus in a number of the other pump tuning parameters (other gates, AC amplitude and source-drain bias) also equal to $ef$. Theoretical models of the electron capture have been successfully applied to modeling the data, and methods of tuning the pump have improved. Operating a pump as a primary standard at the $10^{-7}$ relative uncertainty level requires a more thorough investigation of robustness, as well as the development of an agreed set of guidelines for establishing the operating point of a pump. Increasing the operating frequency of the pumps well above $1$~GHz would be a major breakthrough, enabling robustness studies to be undertaken in a shorter time as well as increasing the useful output current for practical metrological applications.

\begin{acknowledgments}
The authors would like to thank Frank Hohls for supplying raw data from the precision measurements performed at PTB. This research was supported by the UK department for Business, Energy and Industrial Strategy and the EMPIR Joint Research Project `e-SI-Amp' (15SIB08). The European Metrology Programme for Innovation and Research (EMPIR) is co-financed by the Participating States and from the European Union's Horizon 2020 research and innovation programme. A.F. and G.Y. are supported by JSPS KAKENHI Grant Number JP18H05258. M.B. and N.K. are supported by the Korea Research Institute of Standards and Science (KRISS-GP2018-0003). NK was partially supported by the National Research Foundation of Korea (NRF-2016K1A3A7A03951913).
\end{acknowledgments}

\bibliography{SPGrefs_UniversalityPaper}

\end{document}